\pdfoutput=1

\documentclass[11pt,twoside,a4paper,cmspaper,final,collab]{cms-tdr}
\def\svnVersion{475205P}\def\cmsCernNoTag{CERN-EP-2018-089}\def\cmsCernDate{\today}\def\cmsMessage{Published in Physics Letters B as \href{http://dx.doi.org/10.1016/j.physletb.2018.08.057}{\doi{10.1016/j.physletb.2018.08.057.}}}

\begin{document}\cmsNoteHeader{HIG-17-024}

\hyphenation{had-ron-i-za-tion}
\hyphenation{cal-or-i-me-ter}
\hyphenation{de-vices}
\RCS$HeadURL: svn+ssh://svn.cern.ch/reps/tdr2/papers/HIG-17-024/trunk/HIG-17-024.tex $
\RCS$Id: HIG-17-024.tex 468760 2018-07-16 15:02:14Z alverson $

\newlength\cmsTabSkip\setlength{\cmsTabSkip}{1ex}
\newlength\cmsFigWidth
\ifthenelse{\boolean{cms@external}}{\setlength\cmsFigWidth{0.85\columnwidth}}{\setlength\cmsFigWidth{0.4\textwidth}}
\ifthenelse{\boolean{cms@external}}{\providecommand{\cmsLeft}{top\xspace}}{\providecommand{\cmsLeft}{left\xspace}}
\ifthenelse{\boolean{cms@external}}{\providecommand{\cmsRight}{bottom\xspace}}{\providecommand{\cmsRight}{right\xspace}}
\ifthenelse{\boolean{cms@external}}{\providecommand{\cmsLeftUpper}{top\xspace}}{\providecommand{\cmsLeftUpper}{top left\xspace}}
\ifthenelse{\boolean{cms@external}}{\providecommand{\cmsRightUpper}{center\xspace}}{\providecommand{\cmsRightUpper}{top right\xspace}}
\newcommand{\mt}{\ensuremath{m_{\mathrm{T}}}\xspace}
\newcommand{\ma}{\ensuremath{m_{\Pa}}\xspace}
\newcommand{\mbtt}{\ensuremath{m_{\cPqb\Pgt\Pgt}^{\text{vis}}}\xspace}
\newcommand{\mvis}{\ensuremath{m_{\Pgt\Pgt}^{\text{vis}}}\xspace}
\newcommand{\processbbtt}{\ensuremath{{\Ph}\to{\Pa\Pa}\to2\Pgt2{\cPqb}}\xspace}

\cmsNoteHeader{HIG-17-024}

\title{Search for an exotic decay of the Higgs boson to a pair of light pseudoscalars in the final state with two {\cPqb} quarks and two $\Pgt$ leptons in proton-proton collisions at $\sqrt{s}=13\TeV$}

\date{\today}

\abstract{
A search for an exotic decay of the Higgs boson to a pair of light pseudoscalar bosons is performed for the first time in the final state with two {\cPqb} quarks and two $\Pgt$ leptons. The search is motivated in the context of models of physics beyond the standard model (SM), such as two Higgs doublet models extended with a complex scalar singlet (2HDM+S), which include the next-to-minimal supersymmetric SM (NMSSM). The results are based on a data set of proton-proton collisions corresponding to an integrated luminosity of 35.9\fbinv, accumulated by the CMS experiment at the LHC in 2016 at a center-of-mass energy of 13\TeV. Masses of the pseudoscalar boson between 15 and 60\GeV are probed, and no excess of events above the SM expectation is observed. Upper limits between 3 and 12\% are set on the branching fraction $\mathcal{B}(\processbbtt)$ assuming the SM production of the Higgs boson. Upper limits are also set on the branching fraction of the Higgs boson to two light pseudoscalar bosons in different 2HDM+S scenarios. Assuming the SM production cross section for the Higgs boson, the upper limit on this quantity is as low as 20\% for a mass of the pseudoscalar of 40\GeV in the NMSSM.
}

\hypersetup{
pdfauthor={CMS Collaboration},
pdftitle={Search for an exotic decay of the Higgs boson to a pair of light pseudoscalars in the final state with two b quarks and two tau leptons in proton-proton collisions at sqrt(s) = 13 TeV},
pdfsubject={CMS},
pdfkeywords={CMS, physics, Higgs boson, exotic decays, NMSSM, 2HDM+S}}

\maketitle

\section{Introduction}
\label{sec:intro}

Within the standard model (SM), the Brout--Englert--Higgs mechanism~\cite{Englert:1964et,Higgs:1964ia,Higgs:1964pj,Guralnik:1964eu,Higgs:1966ev,Kibble:1967sv} is responsible for electroweak symmetry breaking
and predicts the existence of a scalar particle---the Higgs boson. A particle compatible with the Higgs boson was discovered by the ATLAS and CMS collaborations at the CERN LHC~\cite{Aad:2012tfa,Chatrchyan:2012xdj,CMSobservation125Long}. Measurements of the couplings and properties of this particle leave room for exotic decays to beyond-the-SM particles, with a limit of 34\% on this branching fraction at 95\% confidence level (\CL), using data collected at center-of-mass energies of 7 and 8\TeV~\cite{Khachatryan:2016vau}.

The possible existence of exotic decays of the Higgs boson is well motivated~\cite{Dobrescu:2000jt,Dermisek:2005ar,Dermisek:2006wr,Chang:2008cw,PhysRevD.90.075004,Evans:2017kti}. The decay width of the Higgs boson in the SM is so narrow that a small coupling to a light state could lead to branching fractions of the Higgs boson to that light state of the order of several percent. Additionally, the scalar sector can theoretically serve as a portal that allows matter from a hidden sector to interact with SM particles~\cite{Englert:2011yb}. In general, exotic decays of the Higgs boson are allowed in many models that are consistent with all LHC measurements published so far.

An interesting class of such processes consists of decays to a pair of light pseudoscalar particles, which then decay to pairs of SM particles. This type of process is allowed in various models, including two Higgs doublet models augmented by a scalar singlet (2HDM+S). Seven scalar and pseudoscalar particles are predicted in 2HDM+S. One of them, \Ph, is a scalar that can be compatible with the discovered particle with a mass of 125\GeV, and another, the pseudoscalar \Pa, can be light enough so that $\Ph\to \Pa\Pa$ decays are allowed.

\begin{figure}[hbpt]
\centering
        \includegraphics[width=0.49\textwidth]{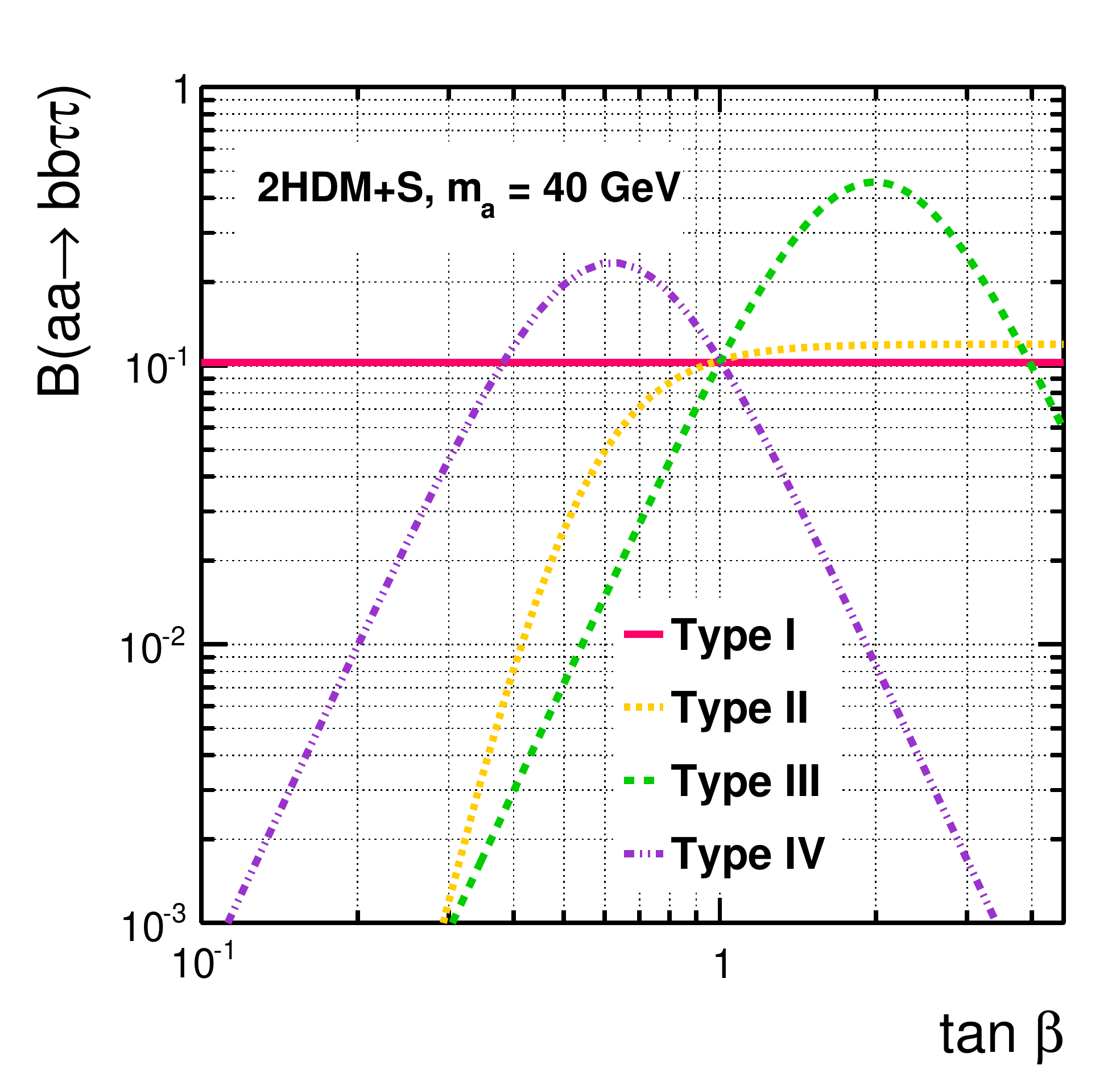}
    \caption{Predicted $\mathcal{B}(\Pa\Pa\to \cPqb\cPqb\Pgt\Pgt)$ for $\ma=40\GeV$ in the different models of 2HDM+S, as a function of $\tan\beta$. The picture is essentially the same for all \ma hypotheses considered in this Letter. The branching fractions are computed following the formulas of Ref.~\cite{PhysRevD.90.075004}.}
    \label{fig:BRbbtt}
\end{figure}

Four types of 2HDM, and by extension 2HDM+S, forbid flavor-changing neutral currents at tree level~\cite{Branco:2011iw}. In type I, all SM particles couple to the first doublet. In type II, up-type quarks couple to the first doublet, whereas leptons and down-type quarks couple to the second doublet. The next-to-minimal supersymmetric SM (NMSSM) is a particular case of 2HDM+S of type II that brings a solution to the $\mu$ problem~\cite{Saul:2010jja}. In type III, quarks couple to the first doublet, and leptons to the second one. Finally, in type IV, leptons and up-type quarks couple to the first doublet, while down-type quarks couple to the second doublet~\cite{PhysRevD.90.075004}. The branching fractions of the light pseudoscalars to pairs of SM particles depend on the type of 2HDM+S, on the pseudoscalar mass \ma, and on $\tan\beta$, defined as the ratio of the vacuum expectation values of the second and first doublets. The value of the branching fraction $\mathcal{B}(\Pa\Pa\to \cPqb\cPqb\Pgt\Pgt)$ is slightly above 10\% in the NMSSM---or more generally in 2HDM+S type II---for $\tan\beta>1$, and can reach up to about 50\% in 2HDM+S type III with $\tan\beta\sim 2$, as shown in Fig.~\ref{fig:BRbbtt}.

Several searches for exotic decays of the Higgs boson to a pair of light short-lived pseudoscalar bosons have been performed by the CMS Collaboration with data collected at a center-of-mass energy of 8\TeV in different final states: $2\Pgm 2\PQb$ for $25.0<\ma<62.5\GeV$~\cite{Khachatryan:2017mnf}, $2\Pgm 2\Pgt$ for $15.0<\ma<62.5\GeV$~\cite{Khachatryan:2017mnf}, $4\Pgt$ for $4<\ma<8\GeV$~\cite{Khachatryan:2015nba} and $5<\ma<15\GeV$~\cite{Khachatryan:2017mnf}, and $4\Pgm$ for $0.25 < \ma < 3.50\GeV$~\cite{Khachatryan:2015wka}. The CMS Collaboration also studied the $2\Pgm 2\Pgt$ final state for $15.0<\ma<62.5\GeV$ at a center-of-mass energy of 13\TeV~\cite{Sirunyan:2018mbx}. The ATLAS Collaboration reported results for the following final states at a center-of-mass energy of 8\TeV: $4\Pgm$, $4\Pe$, and $2\Pe 2\Pgm$ for $15<\ma<60\GeV$~\cite{Aad:2015sva}; $4\Pgg$ for $10<\ma<62\GeV$~\cite{Aad:2015bua}; and $2\Pgm 2\Pgt$ for $3.7<\ma<50.0\GeV$~\cite{Aad:2015oqa}. At a center-of-mass energy of 13\TeV, the ATLAS Collaboration published results for the  $4\cPqb$ decay channel for $20<\ma<60\GeV$~\cite{Aaboud:2016oyb}, and $4\Pgm$, $4\Pe$, and $2\Pe 2\Pgm$ for $1<\ma<60\GeV$~\cite{Aaboud:2018fvk}. The $2\cPqb 2\Pgt$ final state has never been probed so far. This final state benefits from large branching fractions in most models because of the large masses of $\Pgt$ leptons and {\cPqb} quarks with respect to other leptons and quarks. The presence of light leptons originating from the $\Pgt$ decays allows events to be triggered in the dominant gluon fusion production mode.

This Letter reports on the first search with the CMS experiment for exotic decays of the Higgs boson to a pair of light pseudoscalar bosons, in the final state with two $\Pgt$ leptons  and two {\cPqb} quarks. The search focuses on the mass range between 15 and 60\GeV. For low \ma values, between the $\cPqb\cPaqb$ threshold and 15\GeV, the decay products of each of the pseudoscalar bosons become collimated, which would necessitate the use of special reconstruction techniques.

The search is based on proton-proton ($\Pp\Pp$) collision data collected at a center-of-mass energy of 13\TeV and corresponding to an integrated luminosity of 35.9\fbinv. Throughout this Letter, the term $\tauh$ denotes $\Pgt$ leptons decaying hadronically. The $\Pgt\Pgt$ final states studied in this search are $\Pe\Pgm$, $\Pe\tauh$, and $\Pgm\tauh$. Despite its large branching fraction, the $\tauh\tauh$ final state is not considered because the signal acceptance is negligible with the transerse momentum (\pt) thresholds available for the $\tauh\tauh$ triggers. The $\Pe\Pe$ and $\Pgm\Pgm$ final states for the $\Pgt\Pgt$ pair are not considered either, because they have a low branching fraction and suffer from a large contribution of Drell--Yan background events.

\section{The CMS detector}

The central feature of the CMS apparatus is a superconducting solenoid of 6\unit{m} internal diameter, providing a magnetic field of 3.8\unit{T}. Within the solenoid volume, there are a silicon pixel and strip tracker, a lead tungstate crystal electromagnetic calorimeter (ECAL), and a brass and scintillator hadron calorimeter, each composed of a barrel and two endcap sections. Forward calorimeters extend the pseudorapidity coverage provided by the barrel and endcap detectors. Muons are detected in gas-ionization chambers embedded in the steel flux-return yoke outside the solenoid. Events of interest are selected using a two-tiered trigger system~\cite{Khachatryan:2016bia}. A more detailed description of the CMS detector, together with a definition of the coordinate system used and the relevant kinematic variables, can be found in Ref.{}~\cite{Chatrchyan:2008zzk}.

\section{Simulated samples and event reconstruction}

The signal and some of the background processes are modeled with samples of simulated events. The \MGvATNLO~\cite{Alwall:2014hca} 2.3.2 generator is used for the $\processbbtt$ signal process, in gluon fusion ($\cPg\cPg\Ph$), vector boson fusion (VBF), or associated vector boson ($\PW\Ph$, $\PZ\Ph$) processes. These samples are simulated at leading order (LO) in perturbative quantum chromodynamics (QCD) with the MLM jet matching and merging~\cite{Alwall:2007fs}. The $\PZ+\text{jets}$ and $\PW+\text{jets}$ processes are also generated with the \MGvATNLO generator at LO with the MLM jet matching and merging. The $\PZ+\text{jets}$ simulation contains non-resonant Drell--Yan production, with a minimum dilepton mass threshold of 10\GeV. The FxFx merging scheme~\cite{Frederix:2012ps} is used to generate diboson background with the \MGvATNLO generator at next-to-LO (NLO).
The $\ttbar$ and single top quark processes are generated at NLO with the \POWHEG 2.0 and 1.0 generator~\cite{Nason:2004rx,Frixione:2007vw, Alioli:2010xd, Alioli:2010xa, Alioli:2008tz,Frixione:2007nw}.  Backgrounds from SM Higgs boson production are generated at NLO with the \POWHEG 2.0 generator~\cite{Bagnaschi:2011tu}, and the \textsc{minlo hvj}~\cite{Luisoni:2013kna} extension of \POWHEG 2.0 is used for the $\PW\Ph$ and $\PZ\Ph$ simulated samples. The generators are interfaced with \PYTHIA 8.212 ~\cite{Sjostrand:2014zea} to model the parton showering and fragmentation, as well as the decay of the $\Pgt$ leptons. The CUETP8M1 tune~\cite{Khachatryan:2015pea} is chosen for the \PYTHIA parameters that affect the description of the underlying event. The set of parton distribution functions (PDFs) is NLO NNPDF3.0 for NLO samples, and LO NNPDF3.0 for LO samples~\cite{Ball:2014uwa}. The full next-to-NLO (NNLO) plus next-to-next-to-leading logarithmic (NNLL) order calculation~\cite{Beneke:2011mq,Cacciari:2011hy,Baernreuther:2012ws,Czakon:2012pz,Czakon:2012zr,Czakon:2013goa},
performed with the \textsc{Top++} 2.0 program~\cite{Czakon:2011xx}, is used to compute a \ttbar production cross section equal to $832^{+20}_{-29}\,\text{(scale)}\pm 35\,\text{(PDF+\alpS)}\unit{pb}$ setting the top quark mass to 172.5\GeV. This cross section is used to normalize the \ttbar background simulated with \POWHEG.

All simulated samples include additional proton-proton interactions per bunch
crossing, referred to as ``pileup'', obtained by generating concurrent minimum bias collision events using \PYTHIA. The simulated events are reweighted in such a way to have the same pileup distribution as data.
Generated events are processed through a simulation of the CMS detector based on
\GEANTfour~\cite{Agostinelli:2002hh}.

The reconstruction of events relies on the particle-flow (PF) algorithm~\cite{Sirunyan:2017ulk},
which combines information from the CMS subdetectors to identify
and reconstruct the particles emerging from $\Pp\Pp$ collisions:
charged and neutral hadrons, photons, muons, and electrons.
Combinations of these PF objects are used to reconstruct
higher-level objects such as jets, $\tauh$ candidates, and
missing transverse momentum.

Electrons are reconstructed by matching ECAL clusters to tracks in the tracker.  They are then identified with a multivariate analysis (MVA) discriminant that makes use of variables related to the energy deposits in the ECAL, the quality of the track, and the compatibility between the properties of the ECAL clusters and the track that have been matched~\cite{Khachatryan:2015hwa}. The MVA working point chosen in this search has an efficiency of about 80\%. The reconstruction of muon candidates is performed combining the information of the tracker and the muon systems. Muons are then identified on the basis of the track reconstruction quality and on the number of measurements in the tracker and the muon systems~\cite{Chatrchyan:2012xi}.
The relative isolation of electrons and muons is defined as:
\begin{linenomath}
\begin{equation}
I^{\ell} \equiv \frac{\sum_\text{charged}  \PT + \max\left( 0, \sum_\text{neutral}  \PT
                                         - \frac{1}{2} \sum_\text{charged, PU} \PT  \right )}{\PT^{\ell}}.
\label{eq:reconstruction_isolation}
\end{equation}
\end{linenomath}
In this formula, $\sum_\text{charged}  \PT$ is the scalar sum of the
transverse momenta of the charged particles, excluding the lepton itself, associated with
the primary vertex and in a cone around the lepton direction, with size
$\Delta R  = \sqrt{\smash[b]{(\Delta \eta)^2 + (\Delta \phi)^2}} = 0.3$ for electrons, or 0.4 for muons. The sum
$\sum_\text{neutral}  \PT$  represents
a similar quantity for neutral particles. The last term corresponds to the \pt of neutral particles from pileup vertices, which, as estimated from simulation, is equal to approximately half of that of charged hadrons associated with pileup vertices, denoted by $\sum_\text{charged, PU} \PT$. The \pt of the lepton is denoted $\PT^{\ell}$. The azimuthal angle, $\phi$, is measured in radians.

Jets are reconstructed from PF objects with the anti-\kt clustering algorithm implemented
in the \FASTJET library~\cite{Cacciari:2011ma, Cacciari:fastjet2}, using a distance parameter of 0.4. Corrections to the jet energy are applied as a function of the \pt and $\eta$ of the jet~\cite{CMS-JME-10-011}. The jets in this search are required to be separated from the selected electrons, muons, or $\tauh$, by $\Delta R \geq0.5$. Jets that originate from {\cPqb} quarks, called {\cPqb} jets, are identified with the combined secondary vertex (CSVv2) algorithm~\cite{Sirunyan:2017ezt}. The CSVv2 algorithm builds a discriminant from variables related to secondary vertices associated with the jet if any, and from track-based lifetime information. The working point chosen in this search provides an efficiency for {\cPqb} quark jets of approximately 70\%, and a misidentification rate for light-flavor and {\cPqc} quark jets of approximately 1 and 10\%, respectively.

Hadronically decaying $\Pgt$ leptons are reconstructed with the hadrons-plus-strips (HPS) algorithm~\cite{Khachatryan:2015dfa, CMS-PAS-TAU-16-002} as a combination of tracks and energy deposits in strips of the ECAL. The tracks are the signature of the charged hadrons, and the strips that of the neutral pions, which decay to a pair of photons with potential electron-positron conversion. The reconstructed $\tauh$ decay modes are one track, one track plus at least one strip, and three tracks. The rate for jets to be misidentified as $\tauh$ is reduced by applying an MVA discriminator that uses isolation as well as lifetime variables. Its working point has been chosen to have an efficiency of approximately 45\% for a misidentification rate of light-flavor jets of the order of 0.1\%. Additionally, discriminators that reduce the rates with which electrons and muons are misidentified as $\tauh$ are applied. Loose working points with an efficiency above 90\% are chosen because the $\PZ\to\Pe\Pe/\Pgm\Pgm$ background does not contribute much in the region where the signal is expected.

To account for the contribution of undetected particles, such as the neutrinos, the missing transverse momentum, $\ptvecmiss$, is defined as the negative vectorial sum of the transverse momenta of all PF objects reconstructed in the event. The magnitude of this vector is denoted $\ptmiss$. The reconstructed vertex with the largest value of summed physics-object $\pt^2$ is taken to be the primary $\Pp\Pp$ interaction vertex. The physics objects are the objects constructed by a jet finding algorithm~\cite{Cacciari:2008gp,Cacciari:2011ma} applied to all charged tracks associated with the vertex, and the corresponding associated missing transverse momentum.

\section{Event selection}

Events are selected in three different $\Pgt\Pgt$ final states: $\Pe\Pgm$, $\Pe\tauh$, and $\Pgm\tauh$. They are additionally required to contain at least one \cPqb-tagged jet. The dominant backgrounds with these objects in the final state are $\ttbar$ and $\PZ\to\Pgt\Pgt$ production. Another large background consists of events with jets misidentified as $\tauh$, such as $\PW+\text{jets}$ events, the background from SM events composed uniquely of jets produced through the strong interaction, referred to as QCD multijet events, or semileptonic $\ttbar$ events.

All events are required to have at least one \cPqb-tagged jet with $\pt>20\GeV$ and $\abs{\eta}<2.4$. About 90\% of simulated signal events passing this condition have only one such jet, as a result of the typically soft {\cPqb} jet \pt spectrum and of the limited efficiency of the {\cPqb} tagging algorithm. Events in the $\Pe\Pgm$ final state are selected with a trigger that relies on the presence of both an electron and a muon, where the leading lepton has $\pt>23\GeV$ and the subleading one $\pt>12\GeV$ if it is an electron or 8\GeV if it is a muon. In the $\Pe\tauh$ final state, the trigger is based on the presence of an isolated electron with $\pt>25\GeV$, whereas in the $\Pgm\tauh$ final state events are selected with a combination of triggers that require either an isolated muon with $\pt>22\GeV$, or a muon with $\pt>19\GeV$ and a $\tauh$ candidate with $\pt>21\GeV$. During the 2016 data taking period, none of the available triggers that required the presence of both an electron and a $\tauh$ candidate could increase the signal acceptance significantly with respect to the trigger based on the presence of an electron only. Tighter selection criteria are applied at the reconstruction level. The electrons, muons, and $\tauh$ candidates are required to be well identified and isolated~\cite{Khachatryan:2015hwa,Chatrchyan:2012xi,CMS-PAS-TAU-16-002}, to have opposite charge, and to be separated by at least $\Delta R=0.4$ if there is a $\tauh$, or 0.3 otherwise. Table~\ref{tab:baseline} details the \pt, $\eta$, isolation, and identification criteria for the various objects, in the different final states.

\begin{table}
\centering
\topcaption{Baseline selection criteria for objects required in various final states. The numbers given for the \pt thresholds of the electron and muon in the $\Pe\Pgm$ final state correspond to the leading and subleading particles. The \pt threshold for the $\tauh$ candidates is the result of an optimization of the expected exclusion limits of the signal.}\label{tab:baseline}
\begin{tabular}{lccc}
 & $\Pe\Pgm$ & $\Pe\tauh$ & $\Pgm\tauh$ \\
\hline
$\pt(\Pe)$ & ${>}24/13\GeV$ & ${>}26\GeV$ & \NA \\
$\pt(\Pgm)$ & ${>}24/13\GeV$ & \NA & ${>}20\GeV$\\
$\pt(\tauh)$ & \NA & ${>}25\GeV$ & ${>}25\GeV$ \\
$\pt(\cPqb)$ & ${>}20\GeV$ & ${>}20\GeV$ & ${>}20\GeV$ \\
$\abs{\eta(\Pe)}$ & ${<}2.4$ & ${<}2.1$ & \NA \\
$\abs{\eta(\Pgm)}$ & ${<}2.4$ & \NA & ${<}2.1$ \\
$\abs{\eta(\tauh)}$ & \NA & ${<}2.3$ & ${<}2.3$ \\
$\abs{\eta(\cPqb)}$ & ${<}2.4$ & ${<}2.4$ & ${<}2.4$ \\
Isolation ($\Pe$) & ${<}0.10$ & ${<}0.10$ & \NA \\
Isolation ($\Pgm$) & ${<}0.15$ & \NA & ${<}0.15$ \\
Ident. ($\tauh$) & \NA & MVA & MVA \\
\end{tabular}
\end{table}

\begin{figure}[hbt!]
\centering
        \includegraphics[width=0.49\textwidth]{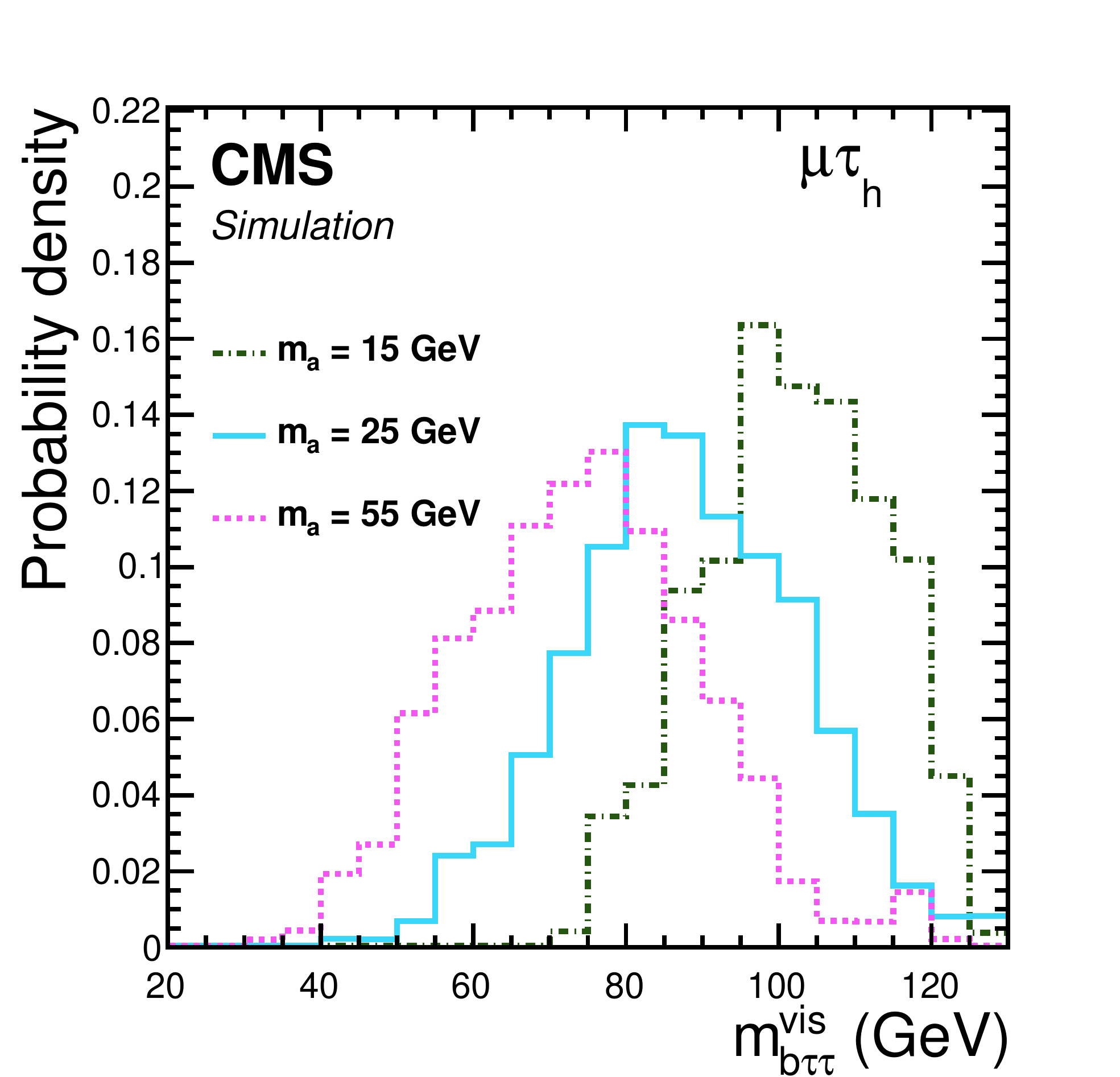}
        \includegraphics[width=0.45\textwidth]{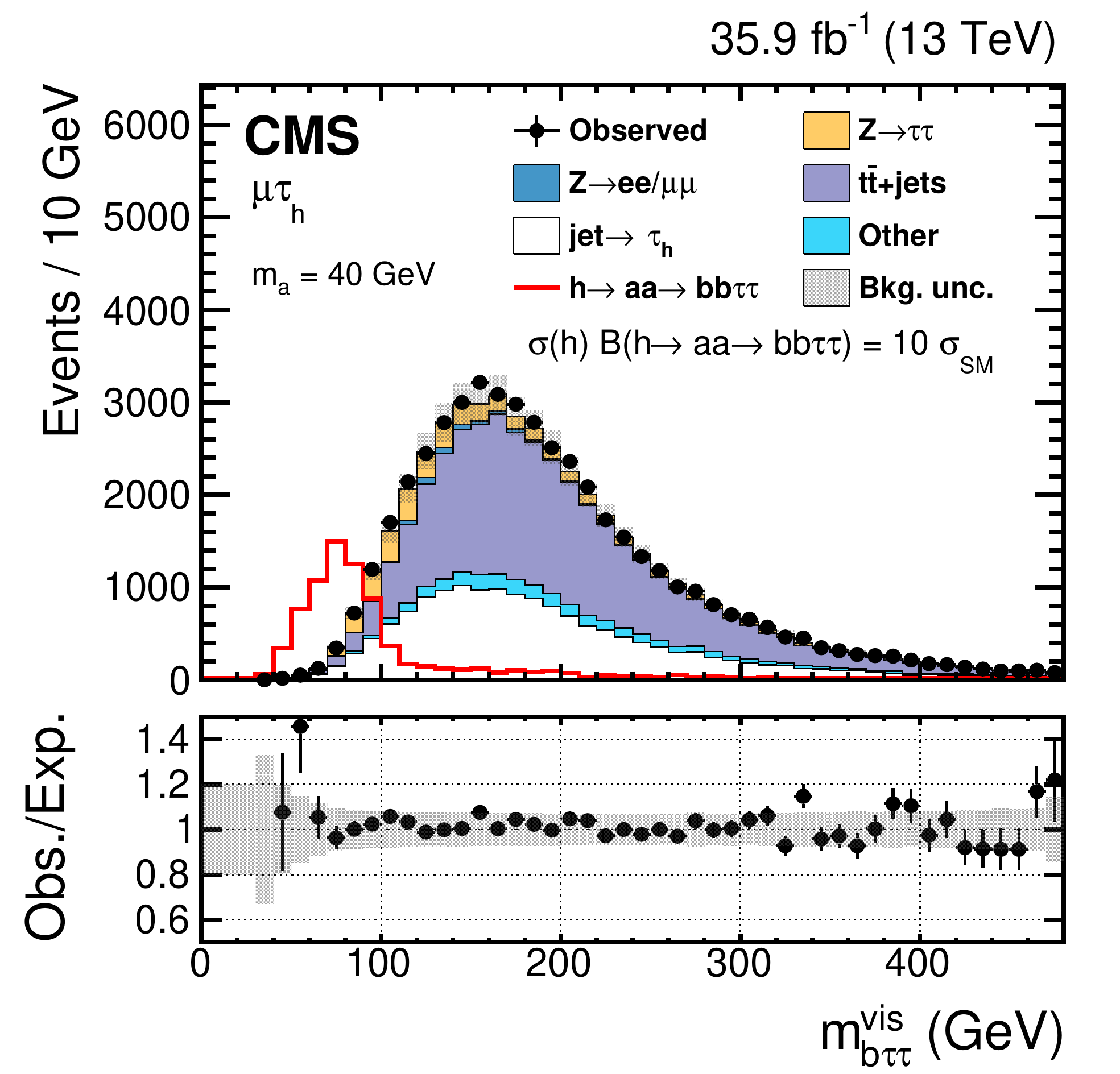}
    \caption{Visible invariant mass of the leptons and the leading {\cPqb} jet, \mbtt, after the baseline selection, in the $\Pgm\tauh$ final state, for the signal with different mass hypotheses (\cmsLeft). Distribution of \mbtt in the $\Pgm\tauh$ final state (\cmsRight). The ``$\text{jet}\to\tauh$" contribution includes all events with a jet misidentified as a $\tauh$ candidate, whereas the rest of background contributions only include events where the reconstructed $\tauh$ corresponds to a $\tauh$, a muon, or an electron, at the generator level. The ``Other" contribution includes events from single top quark, diboson, and SM Higgs boson processes. The signal histogram corresponds to 10 times the SM production cross section for $\cPg\cPg\Ph$, VBF, and V$\Ph$ processes, and assumes $\mathcal{B}(\processbbtt)=100\%$.}
    \label{fig:mbtt}
\end{figure}

To increase the sensitivity of the analysis, events in each final state are separated into four categories with different signal-to-background ratios. The categories are defined on the basis of the invariant mass of the visible decay products of the $\Pgt$ leptons and the \cPqb-tagged jet with the highest \pt, denoted by \mbtt. This variable is typically low for signal events because the three objects originate from a 125\GeV Higgs boson, but it is on average much larger for background events, where the three objects do not originate from a decay of a resonance, as shown in Fig.~\ref{fig:mbtt} for the $\Pgm\tauh$ final state. The thresholds that define the categories depend on the $\Pgt\Pgt$ final state: they are lower in the $\Pe\Pgm$ final state because there are more neutrinos not included in the mass calculation, and they are higher in the $\Pe\tauh$ final state to keep enough events despite the lower signal acceptance related to the electron \pt thresholds. Signal events with $\ma \gtrsim 25\GeV$ contribute mostly to the first two categories, whereas those with $\ma\lesssim 25\GeV$ are concentrated in the second and third categories. This can be explained by the fact that the missing {\cPqb} jet in the mass calculation would be closer to the reconstructed {\cPqb} jet for a signal with lower \ma because of the boost of the pseudoscalar bosons, leading to a larger reconstructed mass. The last category has large background yields; it is useful to constrain the various backgrounds and provides some additional sensitivity for low-\ma signal samples. The results of the search are extracted from a fit of the visible $\Pgt\Pgt$ mass ($\mvis$) distributions in each of the categories, because this is  a resonant distribution for signal events.

\begin{figure}[hbpt]
\centering
        \includegraphics[width=0.38\textwidth]{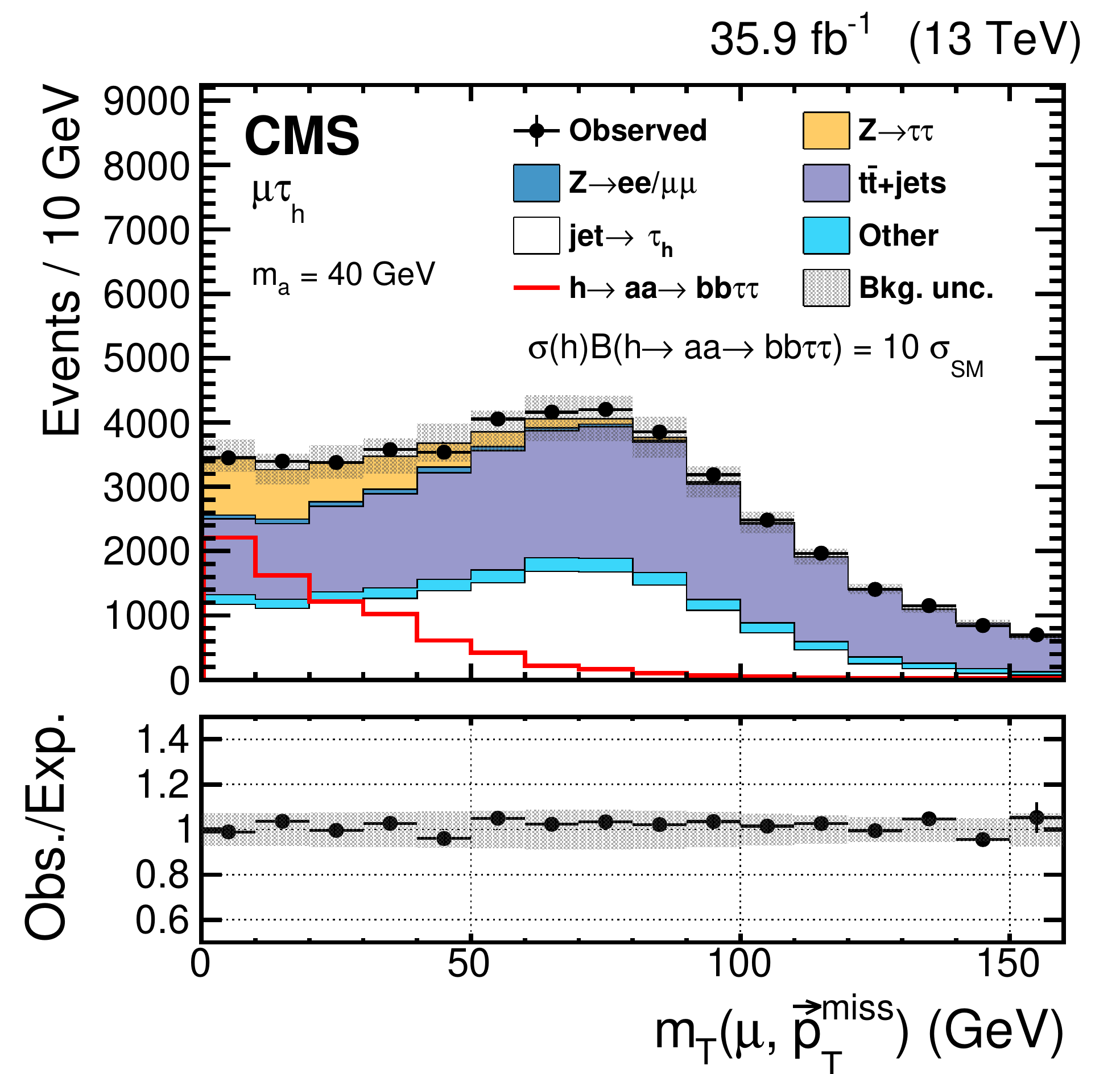}
        \includegraphics[width=0.38\textwidth]{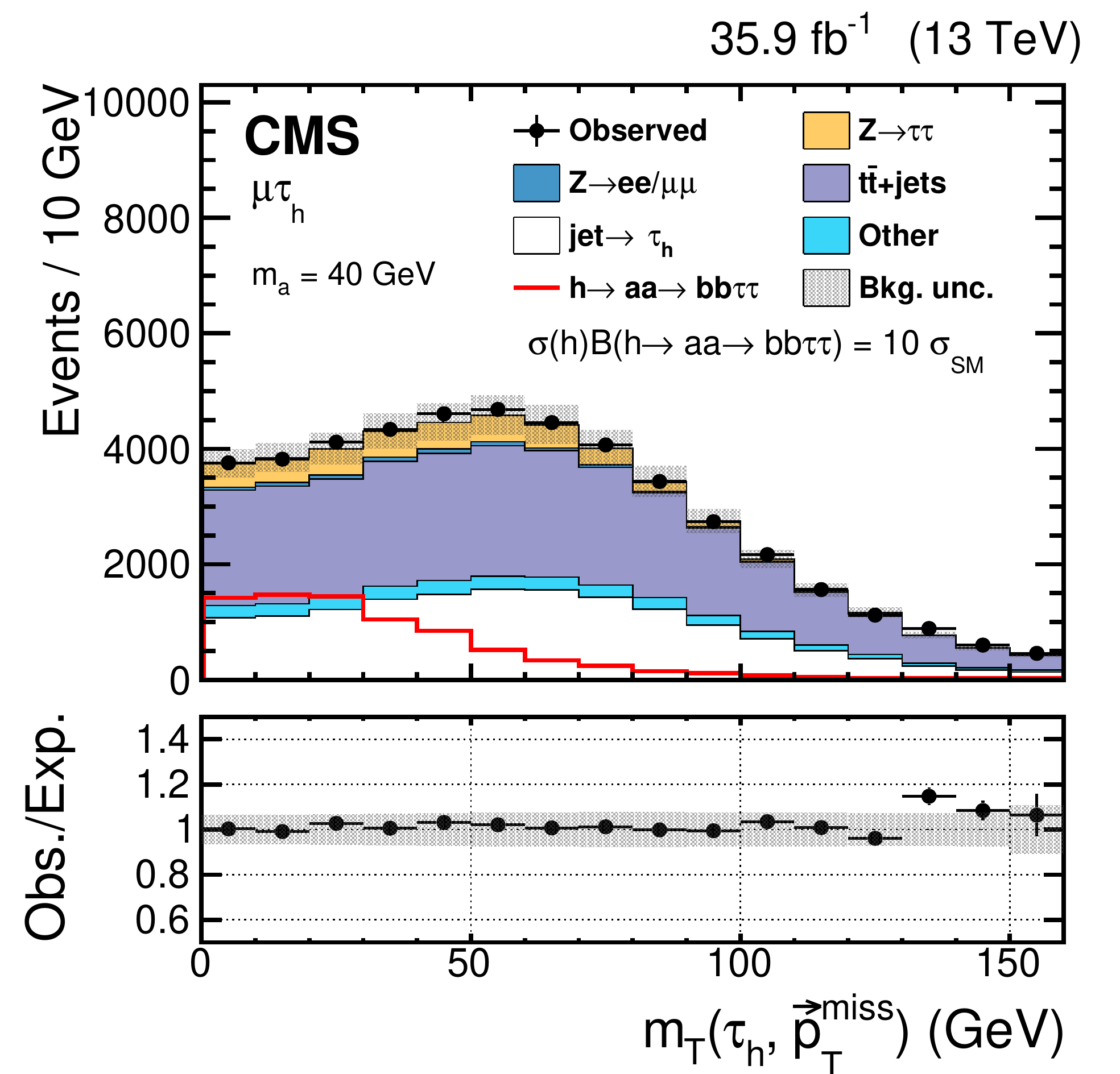}\\
        \includegraphics[width=0.38\textwidth]{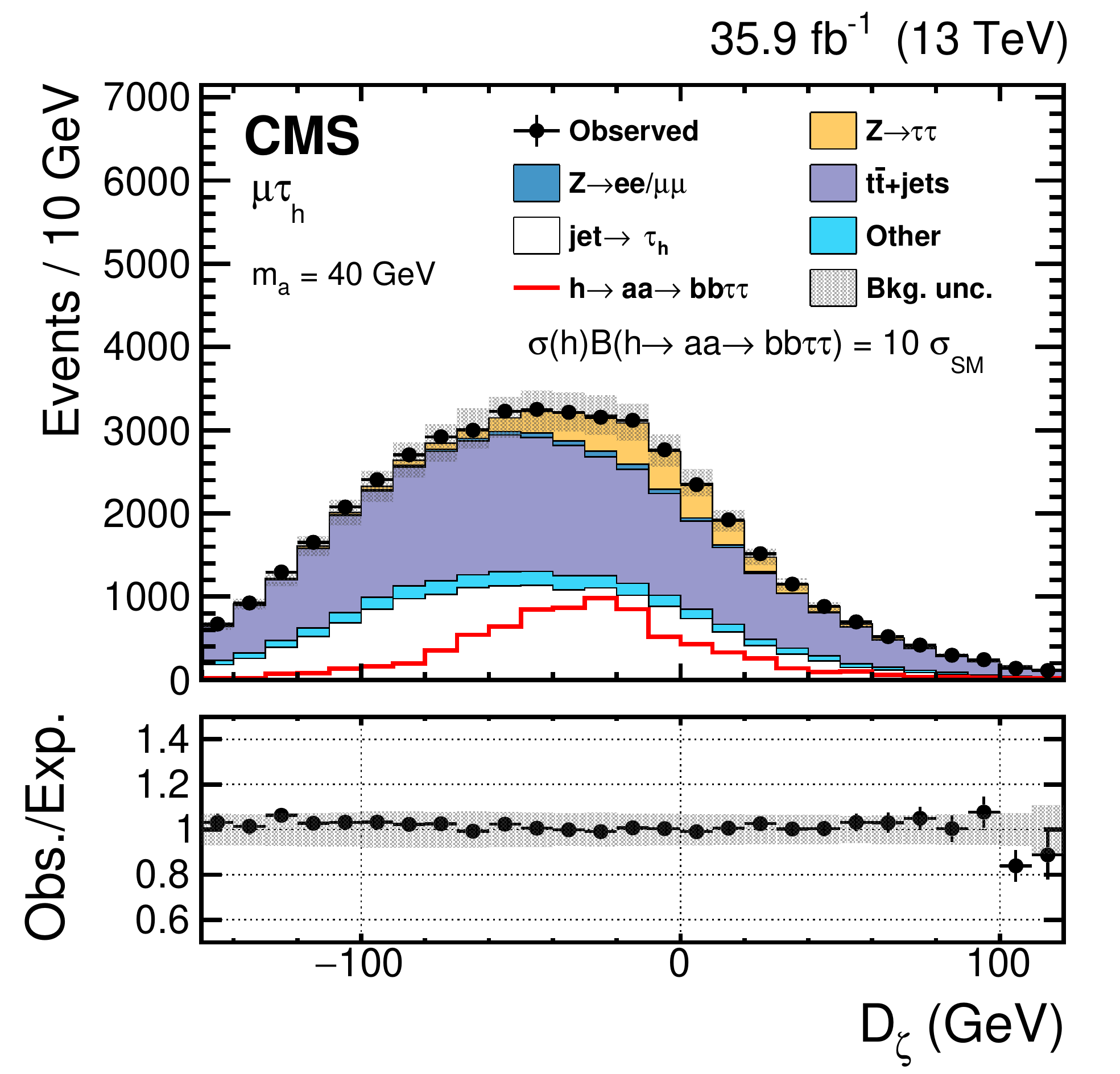}
    \caption{Distributions of $\mt(\Pgm,\ptvecmiss)$ (\cmsLeftUpper), $\mt(\tauh,\ptvecmiss)$ (\cmsRightUpper), and $D_\zeta$ (bottom) in the $\Pgm\tauh$ final state before the $\mbtt$-based categorization. The ``$\text{jet}\to\tauh$" contribution includes all events with a jet misidentified as a $\tauh$ candidate, whereas the rest of background contributions only include events where the reconstructed $\tauh$ corresponds to a $\tauh$, a muon, or an electron, at the generator level. The ``Other" contribution includes events from single top quark, diboson, and SM Higgs boson processes. The signal histogram corresponds to 10 times the SM production cross section for $\cPg\cPg\Ph$, VBF, and V$\Ph$ processes, and assumes $\mathcal{B}(\processbbtt)=100\%$.}
    \label{fig:sel}
\end{figure}

Selection criteria are applied to optimize the expected limits on the product of the signal cross section and branching fraction. The same thresholds would be obtained with an optimization based on the discovery potential. One such criterion is based on the transverse mass of $\ptvecmiss$ and each of the leptons. The transverse mass $\mt$ between a lepton $\ell$ and $\ptvecmiss$ is defined as
\begin{linenomath}
\begin{equation}
\mt(\ell,\ptvecmiss) \equiv \sqrt{\smash[b]{2 \pt^\ell \ptmiss [1-\cos(\Delta\phi)]}} ,
\end{equation}
\end{linenomath}
where $\pt^\ell$ is the transverse momentum of the lepton $\ell$,
and $\Delta\phi$ is the azimuthal angle between the lepton momentum and \ptvecmiss. Selecting events with low $\mt$ strongly reduces the backgrounds from $\PW+\text{jets}$ and $\ttbar$ events, which are characterized by a larger $\ptvecmiss$. In addition, for $\PW+\text{jets}$ events in which the selected lepton comes from the $\PW$ boson decay, $\mt$ has a Jacobian peak near the $\PW$ boson mass. The distributions of $\mt(\Pgm,\ptvecmiss)$ and $\mt(\tauh,\ptvecmiss)$ in the $\Pgm\tauh$ final state before the $\mbtt$-based categorization are shown in Fig.~\ref{fig:sel} (\cmsLeftUpper and \cmsRightUpper).

Another selection criterion is based on the variable $D_\zeta$, which is defined as
\begin{linenomath}
\begin{equation}
D_\zeta \equiv p_\zeta - 0.85 \, p_\zeta^{\text{vis}},
\end{equation}
\end{linenomath}
where $p_\zeta$ is the component of \ptvecmiss along the bisector of the transverse momenta of the two $\Pgt$ candidates and $p_\zeta^{\text{vis}}$ is the sum of the components of the lepton \pt along
the same direction~\cite{Khachatryan:2014wca}. As shown in Fig.~\ref{fig:sel} (bottom), the $\PZ\to\Pgt\Pgt$ background typically has $D_\zeta$ values close to zero because $\ptvecmiss$ is approximately collinear to the $\Pgt\Pgt$ system, whereas the $\ttbar$ background is concentrated at lower $D_\zeta$ values because of typically large $\ptvecmiss$ not aligned with the $\Pgt\Pgt$ system. The signal lies in an intermediate region because $\ptvecmiss$ is approximately aligned with the $\Pgt\Pgt$ system, but $\ptmiss$ is usually small.
The precise criteria for each final state and category are indicated in Table~\ref{tab:sel}.

\begin{table*}
\centering
\topcaption{Optimized selection and categorization in the various final states. The selection criterion $D_\zeta>-30\GeV$ in the $\Pe\Pgm$ final state reduces the large $\ttbar$ background. In the other final states the $\ttbar$ background is less important, and only events with $D_\zeta>0\GeV$ are discarded in one of the categories of the $\Pgm\tauh$ final state to reduce the $\PZ\to\Pgt\Pgt$ background. This selection criterion does not improve the sensitivity in the $\Pe\tauh$ final state because of the lower expected signal and background yields, and is therefore not applied.}\label{tab:sel}
\begin{tabular}{lcccc}
Variable & Category 1 & Category 2 & Category 3 & Category 4 \\
\hline
 & \multicolumn{4}{c}{$\Pe\Pgm$} \\[\cmsTabSkip]
$\mbtt$ & ${<}65\GeV$ & ${\in}[65,80]\GeV$ & ${\in}[80,95]\GeV$ & ${>}95\GeV$ \\
$\mt(\Pe,\ptvecmiss)$ & ${<}40\GeV$ & ${<}40\GeV$ & ${<}40\GeV$ & ${<}40\GeV$  \\
$\mt(\Pgm,\ptvecmiss)$ & ${<}40\GeV$ & ${<}40\GeV$ & ${<}40\GeV$ & ${<}40\GeV$ \\
$D_\zeta$ & ${>}-30\GeV$ & ${>}-30\GeV$ & ${>}-30\GeV$ & ${>}-30\GeV$ \\[\cmsTabSkip]
 & \multicolumn{4}{c}{$\Pe\tauh$} \\[\cmsTabSkip]
$\mbtt$ & ${<}80\GeV$ & ${\in}[80,100]\GeV$ & ${\in}[100,120]\GeV$ & ${>}120\GeV$ \\
$\mt(\Pe,\ptvecmiss)$ & ${<}40\GeV$ & ${<}50\GeV$ & ${<}50\GeV$ & ${<}40\GeV$  \\
$\mt(\tauh,\ptvecmiss)$ & ${<}60\GeV$ & ${<}60\GeV$ & ${<}60\GeV$ & ${<}60\GeV$ \\[\cmsTabSkip]
 & \multicolumn{4}{c}{$\Pgm\tauh$} \\[\cmsTabSkip]
$\mbtt$ & ${<}75\GeV$ & ${\in}[75,95]\GeV$ & ${\in}[95,115]\GeV$ & ${>}115\GeV$ \\
$\mt(\Pgm,\ptvecmiss)$ & ${<}40\GeV$ & ${<}50\GeV$ & ${<}50\GeV$ & ${<}40\GeV$  \\
$\mt(\tauh,\ptvecmiss)$ & ${<}60\GeV$ & ${<}60\GeV$ & ${<}60\GeV$ & ${<}60\GeV$ \\
$D_\zeta$ & \NA & ${<}0\GeV$ & \NA  & \NA \\
\end{tabular}
\end{table*}

\section{Background estimation}

The $\PZ\to\ell\ell$ background is estimated from simulation. The distributions of the \pt of the dilepton system and the visible invariant mass between the leptons and the leading {\cPqb} jet are reweighted with corrections derived using data from a region enriched in $\PZ\to\Pgm\Pgm$ + $\geq 1$ {\cPqb} events. The simulation is separated between $\PZ\to\Pgt\Pgt$, where the reconstructed $\Pgt$ candidates correspond to $\Pgt$ leptons at generator level, and $\PZ\to\Pe\Pe/\Pgm\Pgm$ decays, where at least one electron or muon is misidentified as a $\tauh$ candidate.

Backgrounds with a jet misidentified as a $\tauh$ candidate are estimated from data. They consist mostly of $\PW+\text{jets}$ and QCD multijet events, as well as the fraction of $\ttbar$, diboson, and single top quark production where the reconstructed $\tauh$ candidate comes from a jet. The probabilities for jets to be misidentified as $\tauh$ candidates, denoted $f$, are estimated from $\PZ\to\Pgm\Pgm+\text{jets}$ events in data. They are parameterized with Landau distributions as a function of the \pt of the $\tauh$ candidate, separately for every reconstructed $\tauh$ decay mode. Events that pass all the selection criteria, except that the $\tauh$ candidate fails the isolation condition, are reweighted with a weight $f/(1-f)$ to estimate the contribution of events with jets in the signal region. The contribution of events with genuine electrons, muons, or $\tauh$ candidates in the control region is estimated from simulation and subtracted from data.

In the $\Pe\Pgm$ final state, the small $\PW+\text{jets}$ background is estimated from simulation~\cite{Sirunyan:2017khh}. Such events typically have a genuine lepton coming from the $\PW$ boson decay and a jet misidentified as the other lepton. The QCD multijet background, which also contains jets misidentified as leptons, is estimated from data. Its normalization corresponds to the difference between the data and the sum of all the other backgrounds in a so-called same-sign region where the $\Pgt$ candidates have the same sign. A smooth distribution is obtained by additionally relaxing the isolation conditions of both leptons. A correction that is extracted from data is applied to extrapolate the normalization obtained in the same-sign region to the signal region.

Other processes, including diboson, $\ttbar$, and single top quark production without jet misidentified as a $\tauh$ candidate, as well as SM Higgs boson processes in various production and decay modes, are estimated from simulation. The $\ttbar$ production is a major background, especially in the $\Pe\Pgm$ final state. The $\ttbar$ simulation models the variables used in this analysis well, as it has been verified in a control region in the $\Pe\Pgm$ final state where no selection criterion is applied on $\mt(\Pe,\ptvecmiss)$ or $\mt(\Pgm,\ptvecmiss)$, and where the $D_\zeta$ selection criterion is inverted.

In the $\Pe\tauh$ and $\Pgm\tauh$ final states, where all backgrounds with a jet misidentified as a $\tauh$ candidate are estimated from data, simulated events with a reconstructed $\tauh$ that is not matched to an electron, a muon, or a $\tauh$ at the generator level are discarded to avoid double counting. Approximately 30\% of simulated $\ttbar$ events after the selection have a reconstructed $\tauh$ that is not matched to an electron, a muon, or a $\tauh$ at the generator level.

\section{Fit method and systematic uncertainties}

The search for an excess of signal events over the expected background involves a global binned maximum likelihood fit based on the $\mvis$ distributions in the different channels and categories. The statistical uncertainty largely dominates over systematic uncertainties in this search. The systematic uncertainties are represented by nuisance parameters that are varied in the fit according to their probability density functions.
A log-normal probability density function is assumed for the nuisance parameters that affect the event yields of the various background and signal contributions, whereas systematic uncertainties that affect the distributions are represented by nuisance parameters whose variation results in a continuous perturbation of the spectrum~\cite{Conway-PhyStat} and which are assumed to have a Gaussian probability density function.

To take into account the limited size of simulated samples and of data in the control regions used to estimate some of the background processes, statistical uncertainties in individual bins of the $\mvis$ distributions are considered as Poissonian nuisance parameters. The uncertainty can be as large as 40\% for some bins in the low-$\mbtt$ categories. The combined effect of all these uncertainties is the dominant systematic uncertainty in this search.

The uncertainties in the jet energy scale~\cite{CMS-JME-10-011} affect the overall yields of processes estimated from simulation, as well as their relative contribution to the different categories because the categorization is based on the value of $\mbtt$ for each event. They are functions of the jet \pt and $\eta$. The $\ptvecmiss$ is recomputed for each variation of the jet energy scale. The uncertainty in $\ptvecmiss$ related to the measurement of the energy that is not clustered in jets~\cite{Khachatryan:2014gga} is evaluated event-by-event, and is also considered as a shape uncertainty.

Corrections for the efficiency of the identification of electrons, muons, and $\tauh$ candidates are derived from data using tag-and-probe methods~\cite{CMS:2011aa}, and are applied to simulated events as a function of the lepton \pt and $\eta$. Uncertainties related to these corrections amount to 2\% for electrons, 2\% for muons, and 5\% for $\tauh$ candidates. Additionally, events with an electron or muon misidentified as a $\tauh$ candidate have a yield uncertainty of 5\%. Trigger scale factors are also estimated with tag-and-probe methods and their corresponding uncertainties in the yields of simulated processes are 1\%.

The energy scale of $\tauh$ candidates is corrected for each reconstructed decay mode, and the uncertainty of 1.2\% for each single decay mode is considered as a shape uncertainty. Uncertainties in the energy scales of electrons and muons are also included as shape uncertainties.

Corrections to the efficiency for identifying a {\cPqb} quark jet as a {\cPqb} jet, as well as for mistagging a jet originating from a different flavor, are applied to simulated events on the basis of the generated flavor of the jets. The uncertainties in the scale factors depend on the \pt of the jet and are therefore considered as shape uncertainties. They amount to 1.5\% for a jet originating from a {\cPqb} quark, 5\% from a {\cPqc} quark, and 10\% from a light-flavor parton~\cite{Sirunyan:2017ezt}.

The uncertainty in the yield of the backgrounds with jets misidentified as $\tauh$ candidates accounts for possibly different misidentification rates in $\PZ+\text{jets}$ events (where the misidentification rates are measured), and in $\PW+\text{jets}$ and QCD multijet events (which dominate the constitution of the reducible background in the signal region), and for differences between data and predicted backgrounds observed in a region enriched in reducible background events by inverting the charge requirement on the $\Pgt$ candidates and removing the $\mt$ and $D_\zeta$ selection criteria. This uncertainty amounts to 20\%, and is constrained to about 7\% after the maximum likelihood fit because of the large number of events contributing to the last $\mbtt$ category. Uncertainties in the parameterization of the misidentification probability of jets as a function of \pt result in shape uncertainties for the backgrounds with jets misidentified as $\tauh$ candidates.

The uncertainty in the yield of the QCD multijet background in the $\Pe\Pgm$ final state is 20\%; the value comes from the uncertainty in the extrapolation factor from the same-sign region to the opposite-sign region. The uncertainty in the $\PW+\text{jets}$ background in this channel also amounts to 20\%, and accounts for a potential mismodeling in simulation of the misidentification rate of jets as electrons or muons.

The theoretical yield uncertainty of the $\ttbar$ background is related to the PDF uncertainty and to the uncertainty associated to the strong coupling constant \alpS in the full NNLO plus NNLL order calculation of the cross section; it amounts to about 4\%. The yield uncertainties for other backgrounds estimated from simulation are taken from recent CMS measurements: 6\% for diboson processes~\cite{Sirunyan:2017zjc}, 13\% for single top quark processes~\cite{Sirunyan:2016cdg}, and 7\% for $\PZ+\text{jets}$ events with at least one \cPqb-tagged jet in the final state~\cite{Khachatryan:2016iob}. The uncertainty in the correction of the dilepton \pt distribution for Drell--Yan events is equal to 10\% of the size of the correction itself. The uncertainty in the correction of the $\mbtt$ distribution is equal to the correction itself, and considered as a shape uncertainty. Uncertainties in the production cross sections and branching fractions for SM Higgs boson processes are taken from Ref.~{}\cite{deFlorian:2016spz}. The uncertainty in the integrated luminosity amounts to 2.5\%~\cite{CMS-PAS-LUM-17-001}.

\section{Results}
\label{sec:results}

\begin{figure*}[hbpt]
\centering
        \includegraphics[width=0.49\textwidth]{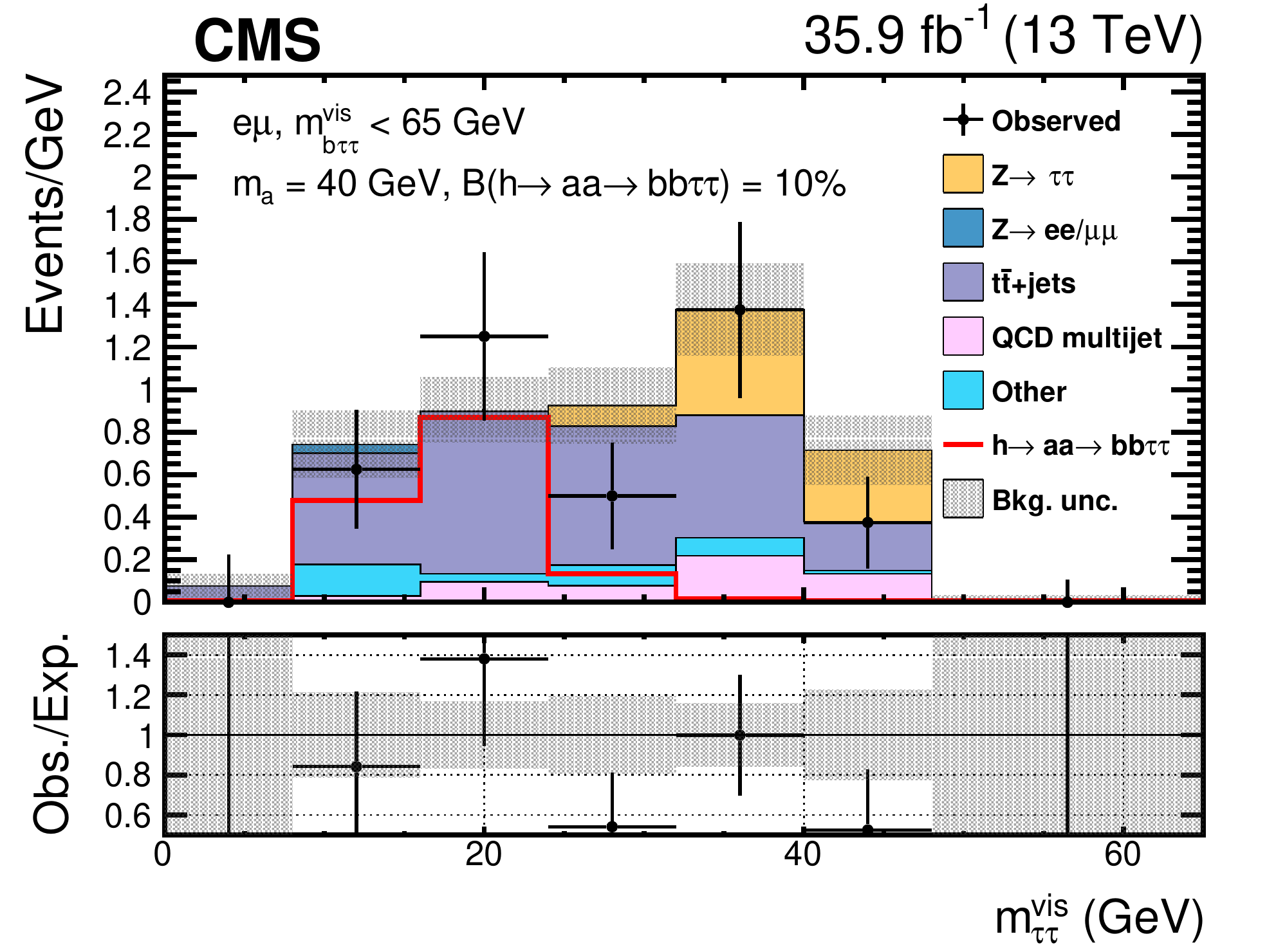}
        \includegraphics[width=0.49\textwidth]{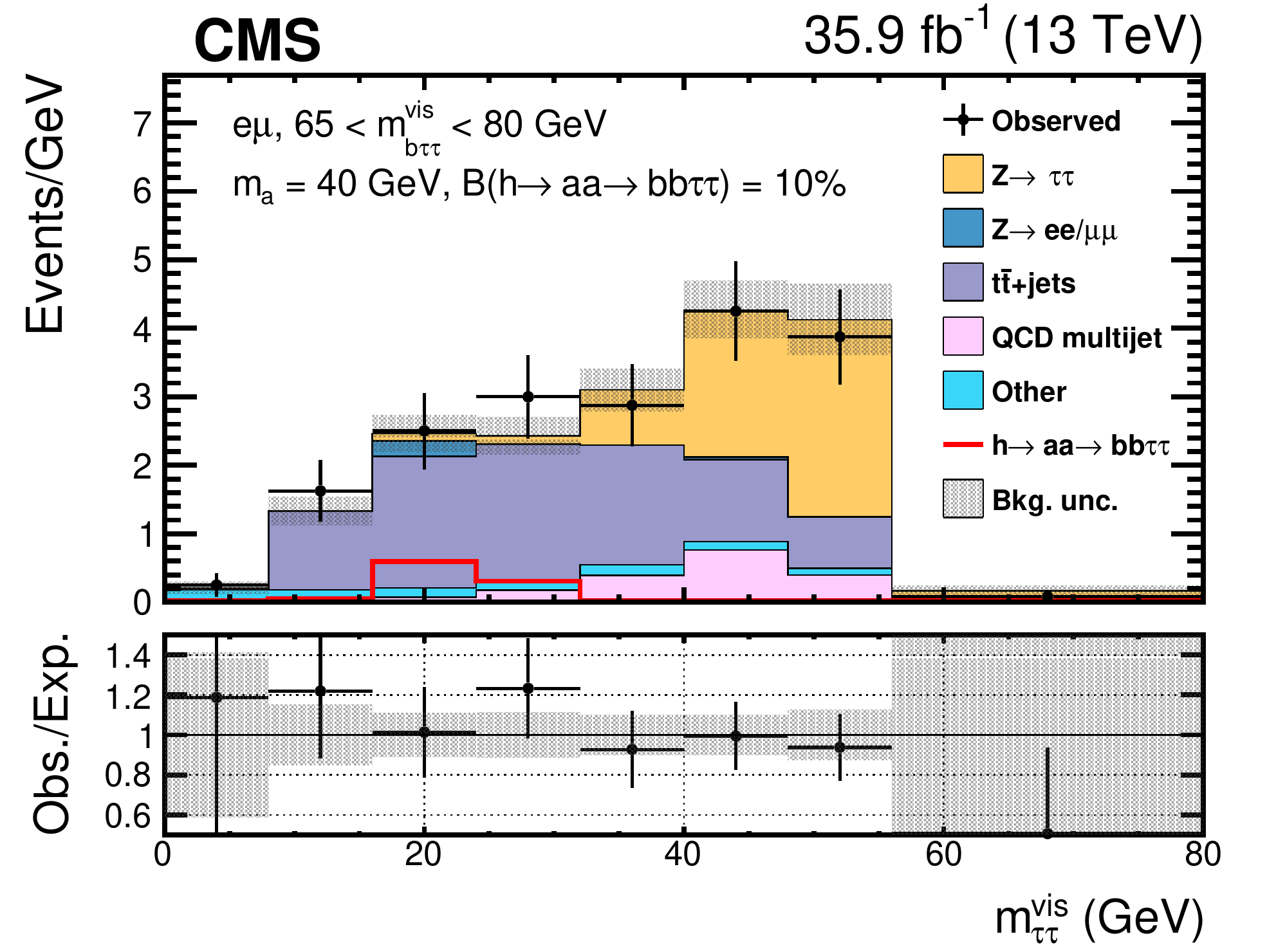}\\
        \includegraphics[width=0.49\textwidth]{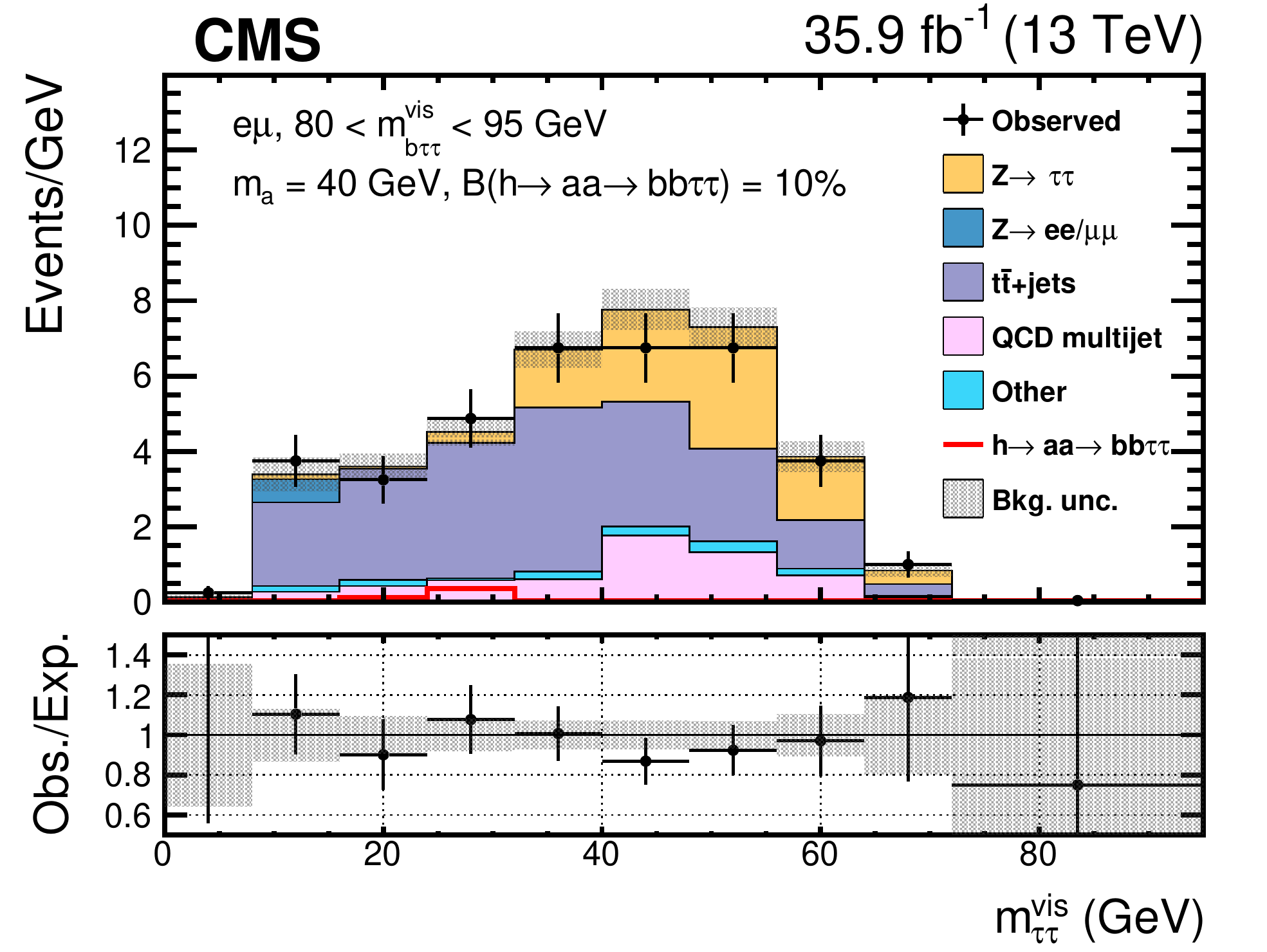}
        \includegraphics[width=0.49\textwidth]{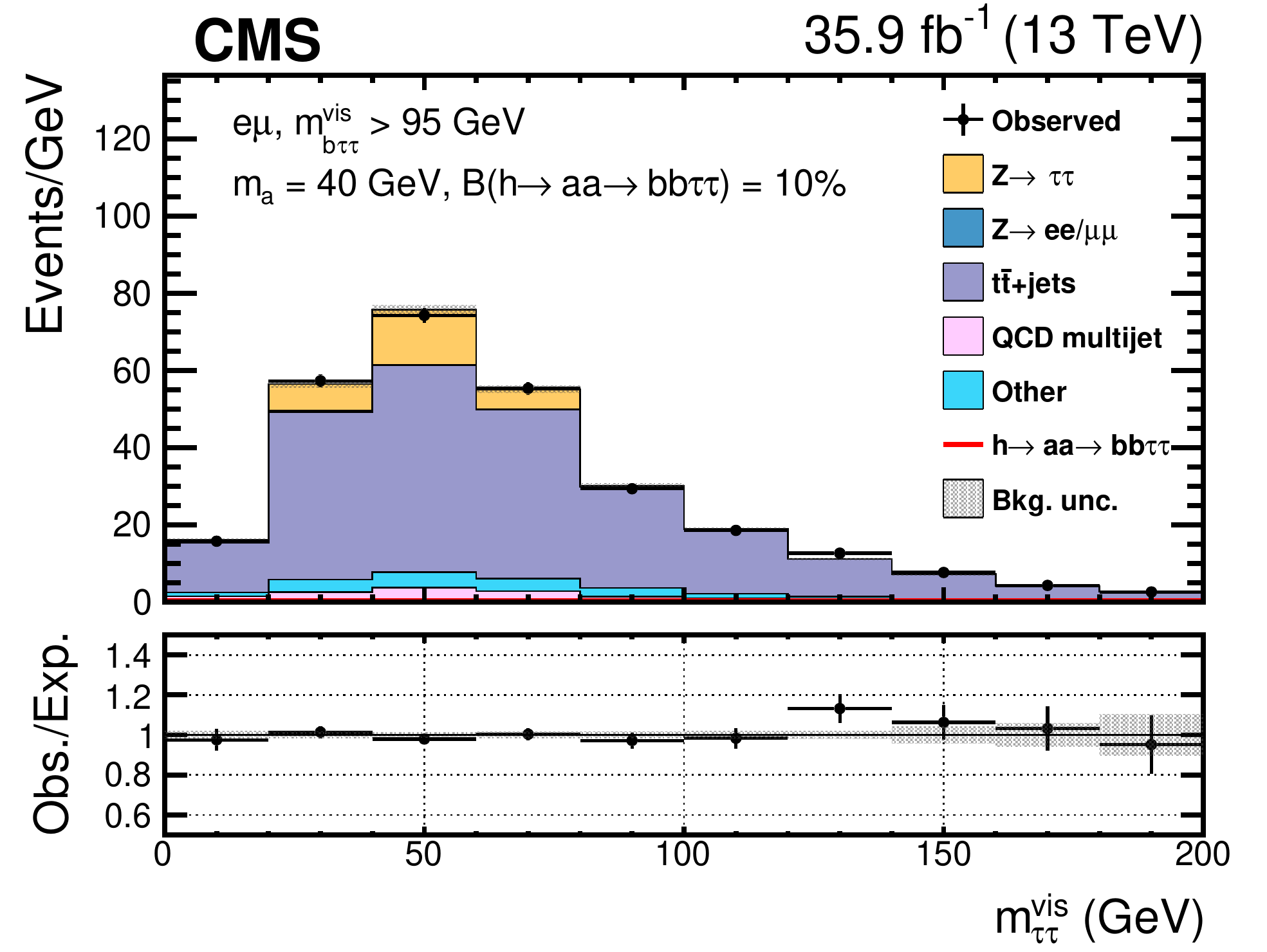}
    \caption{Distributions of $\mvis$ in the four categories of the $\Pe\Pgm$ channel. The ``Other" contribution includes events from single top quark, diboson, SM Higgs boson, and $\PW+\text{jets}$ productions. The signal histogram corresponds to the SM production cross section for $\cPg\cPg\Ph$, VBF, and V$\Ph$ processes, and assumes $\mathcal{B}(\processbbtt)=10\%$. The normalizations of the predicted background distributions correspond to the result of the global fit.}
    \label{fig:em_mtt}
\end{figure*}

\begin{figure*}[hbpt]
\centering
        \includegraphics[width=0.49\textwidth]{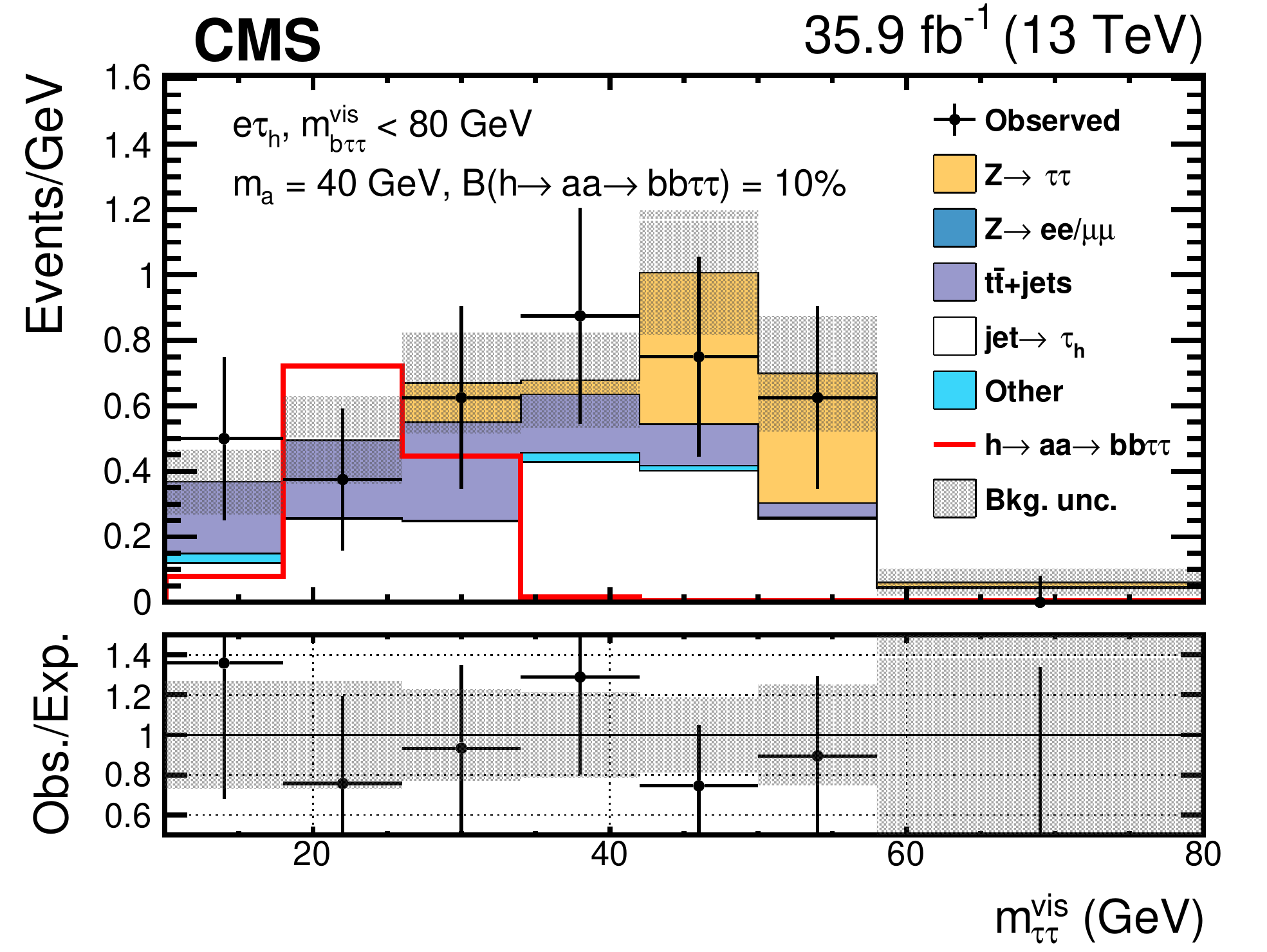}
        \includegraphics[width=0.49\textwidth]{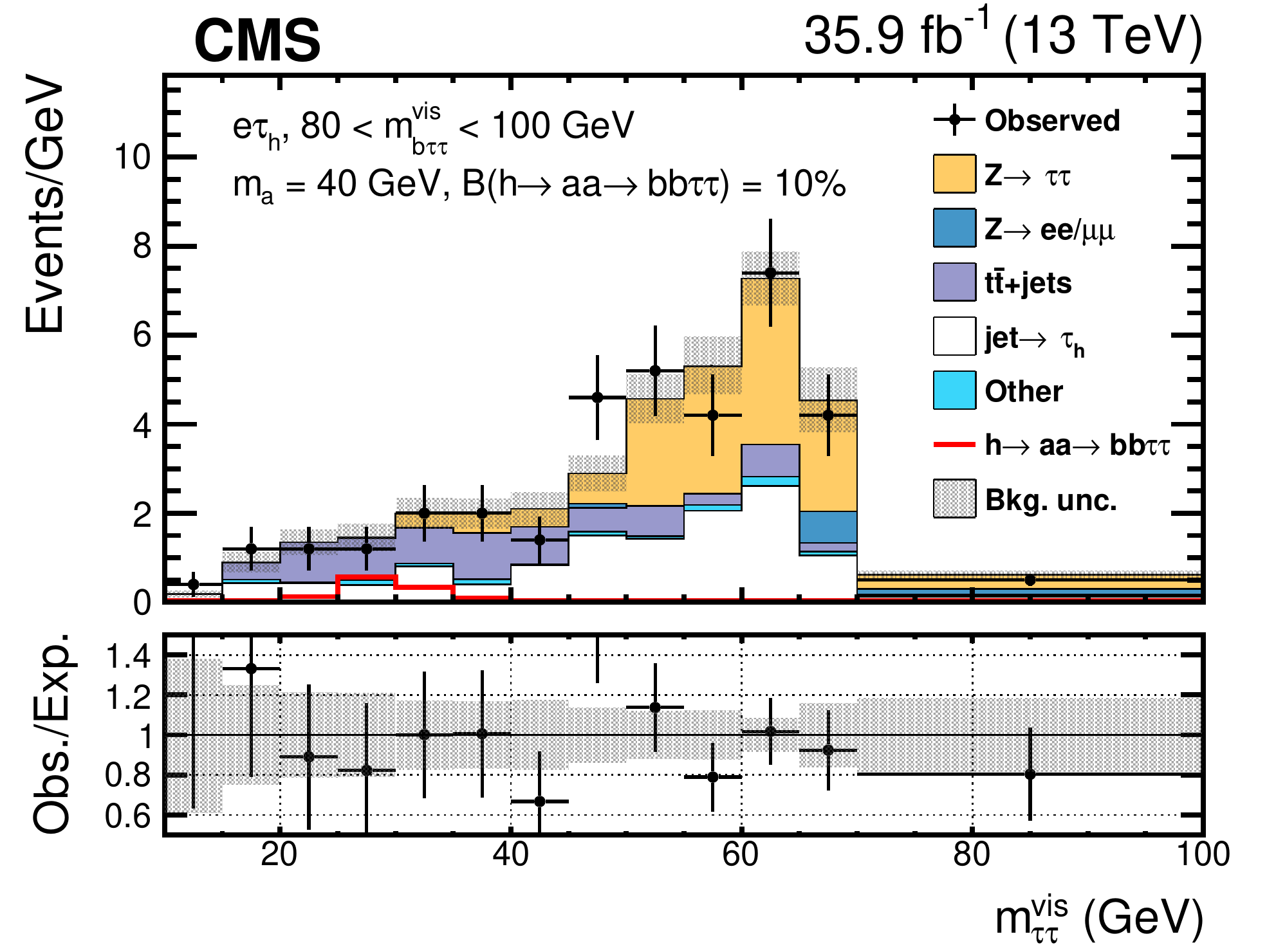}\\
        \includegraphics[width=0.49\textwidth]{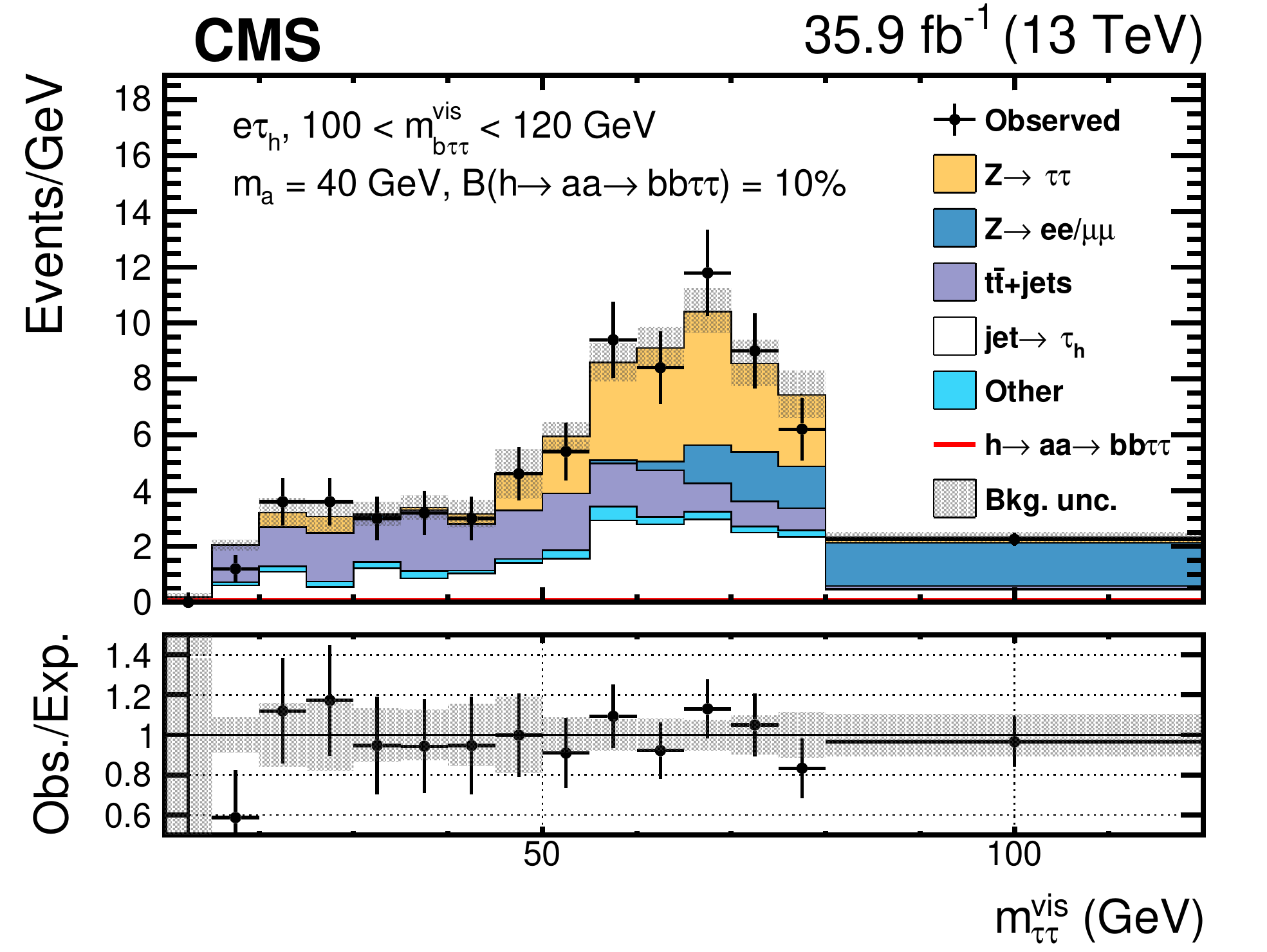}
        \includegraphics[width=0.49\textwidth]{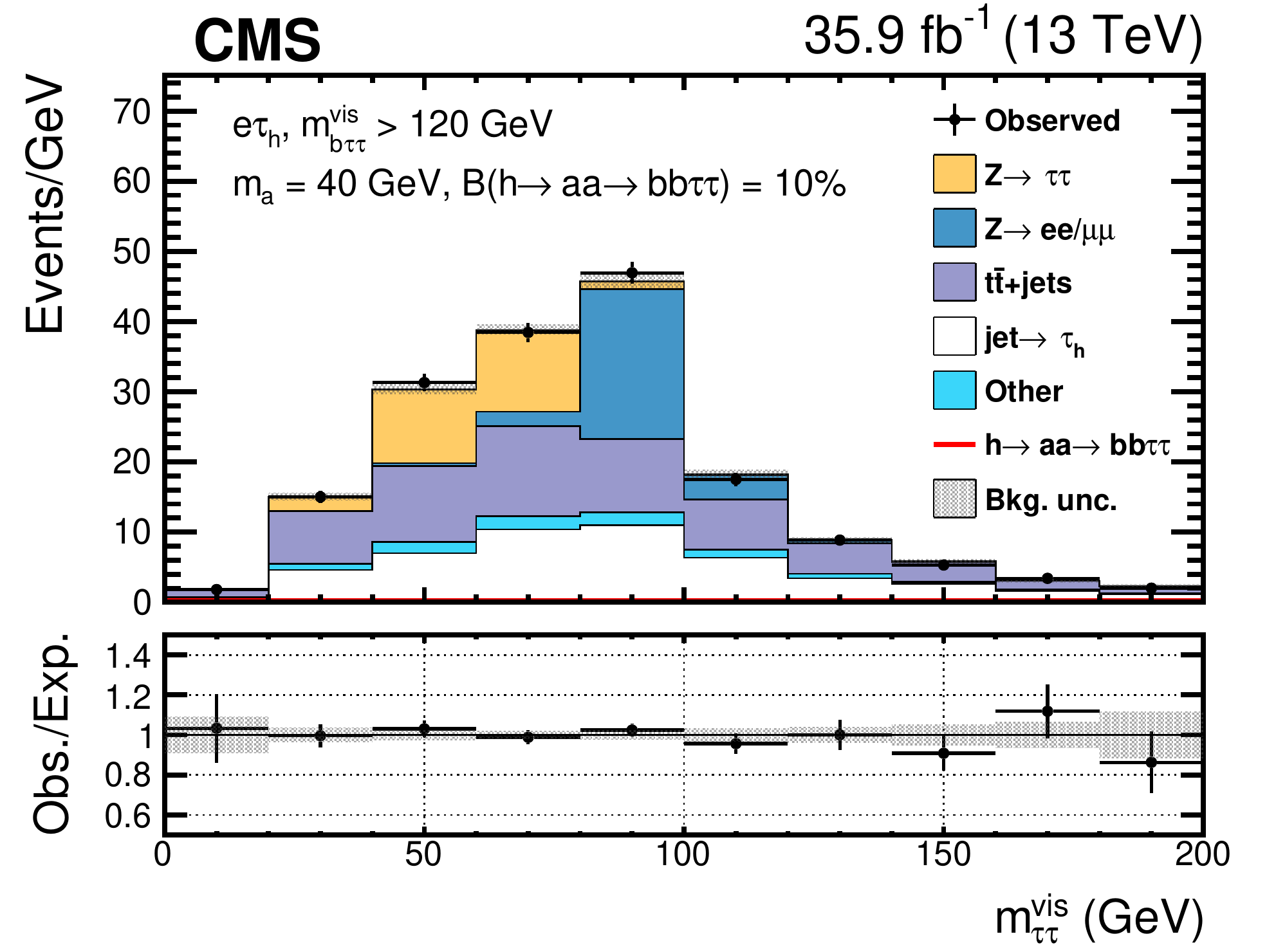}
    \caption{Distributions of $\mvis$ in the four categories of the $\Pe\tauh$ channel. The ``$\text{jet}\to\tauh$" contribution includes all events with a jet misidentified as a $\tauh$ candidate, whereas the rest of background contributions only include events where the reconstructed $\tauh$ corresponds to a $\tauh$, a muon, or an electron, at the generator level. The ``Other" contribution includes events from single top quark, diboson, and SM Higgs boson processes. The signal histogram corresponds to the SM production cross section for $\cPg\cPg\Ph$, VBF, and V$\Ph$ processes, and assumes $\mathcal{B}(\processbbtt)=10\%$. The normalizations of the predicted background distributions correspond to the result of the global fit.}
    \label{fig:et_mtt}
\end{figure*}

\begin{figure*}[hbpt]
\centering
        \includegraphics[width=0.49\textwidth]{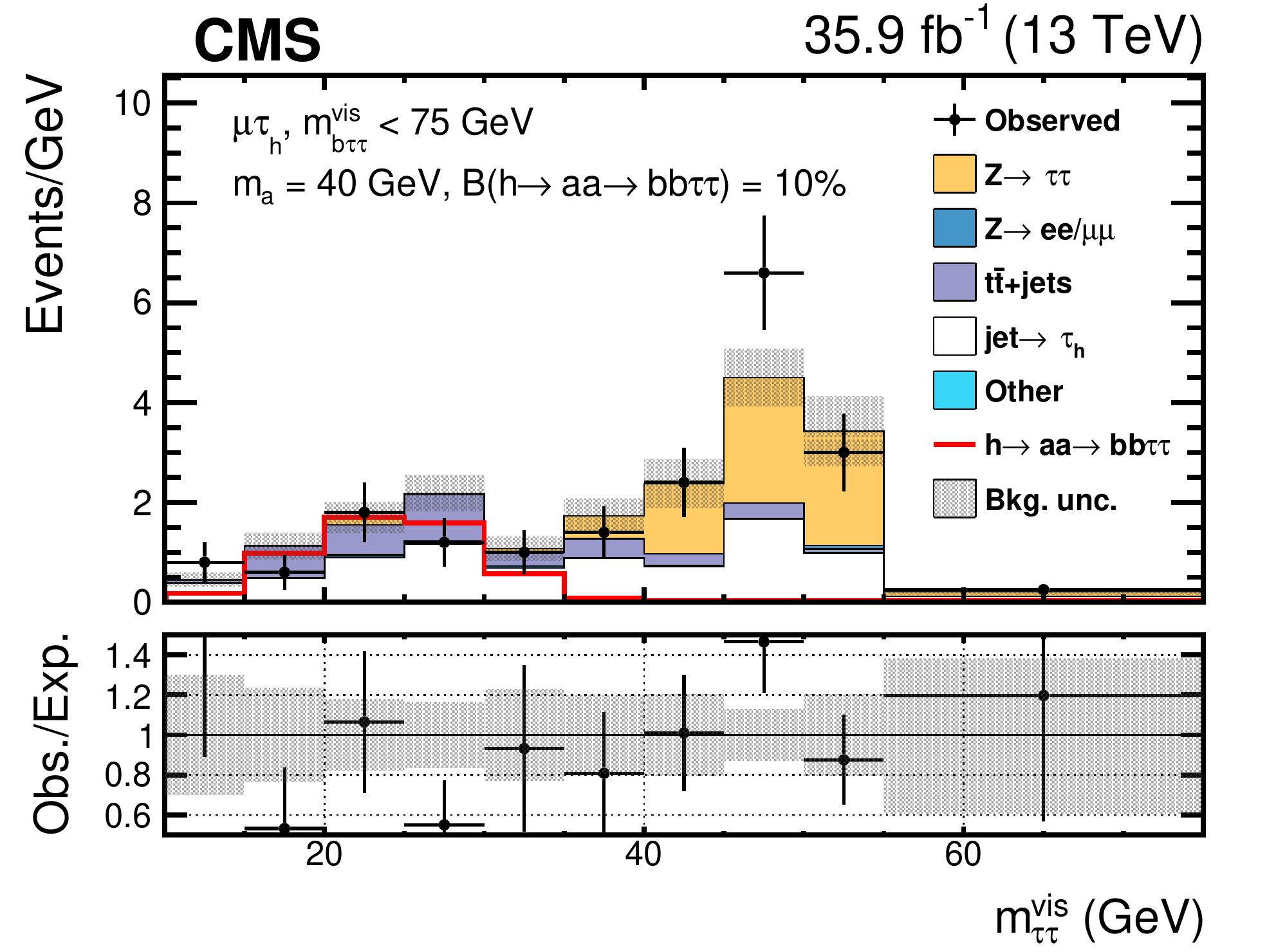}
        \includegraphics[width=0.49\textwidth]{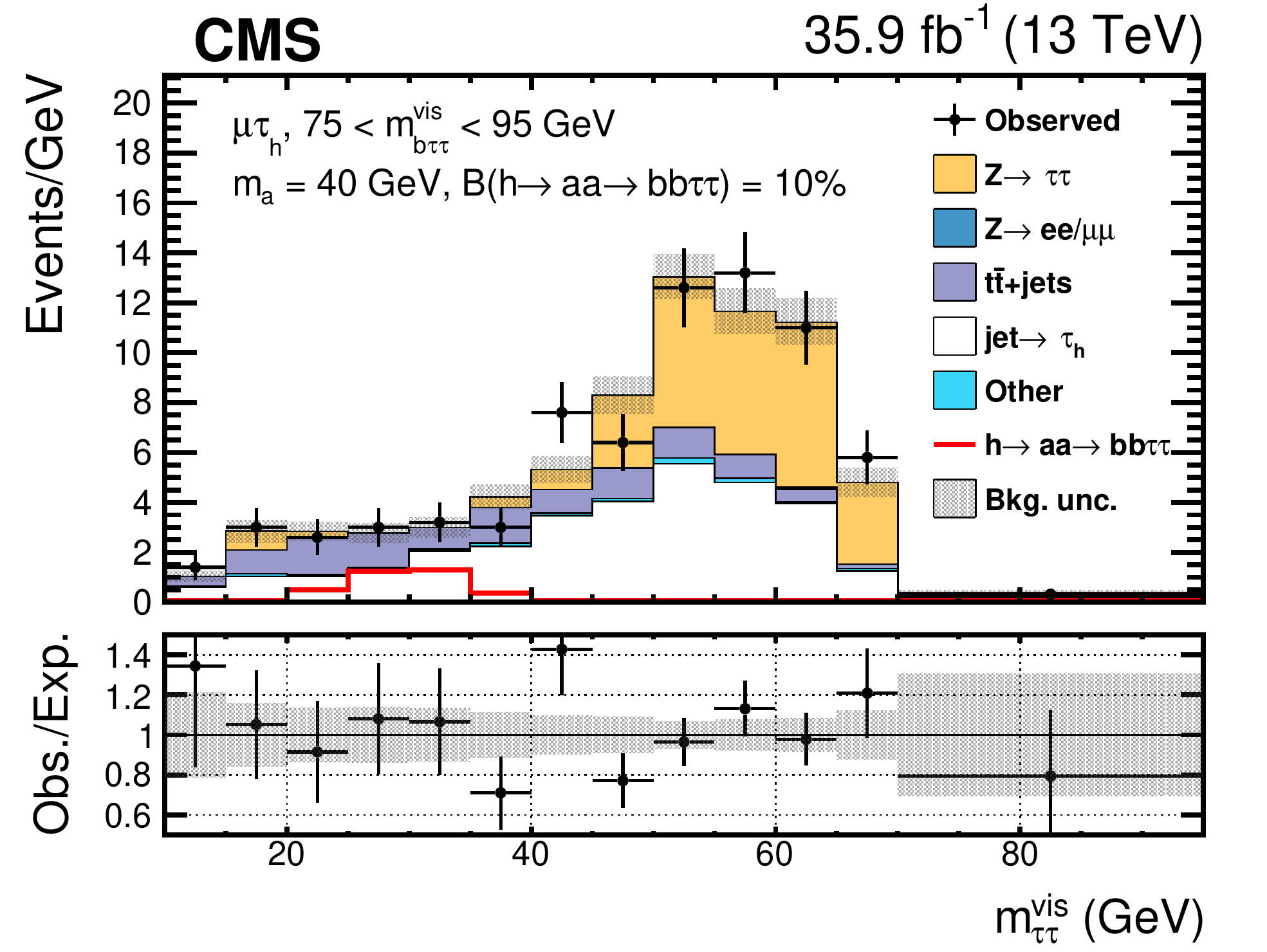}\\
        \includegraphics[width=0.49\textwidth]{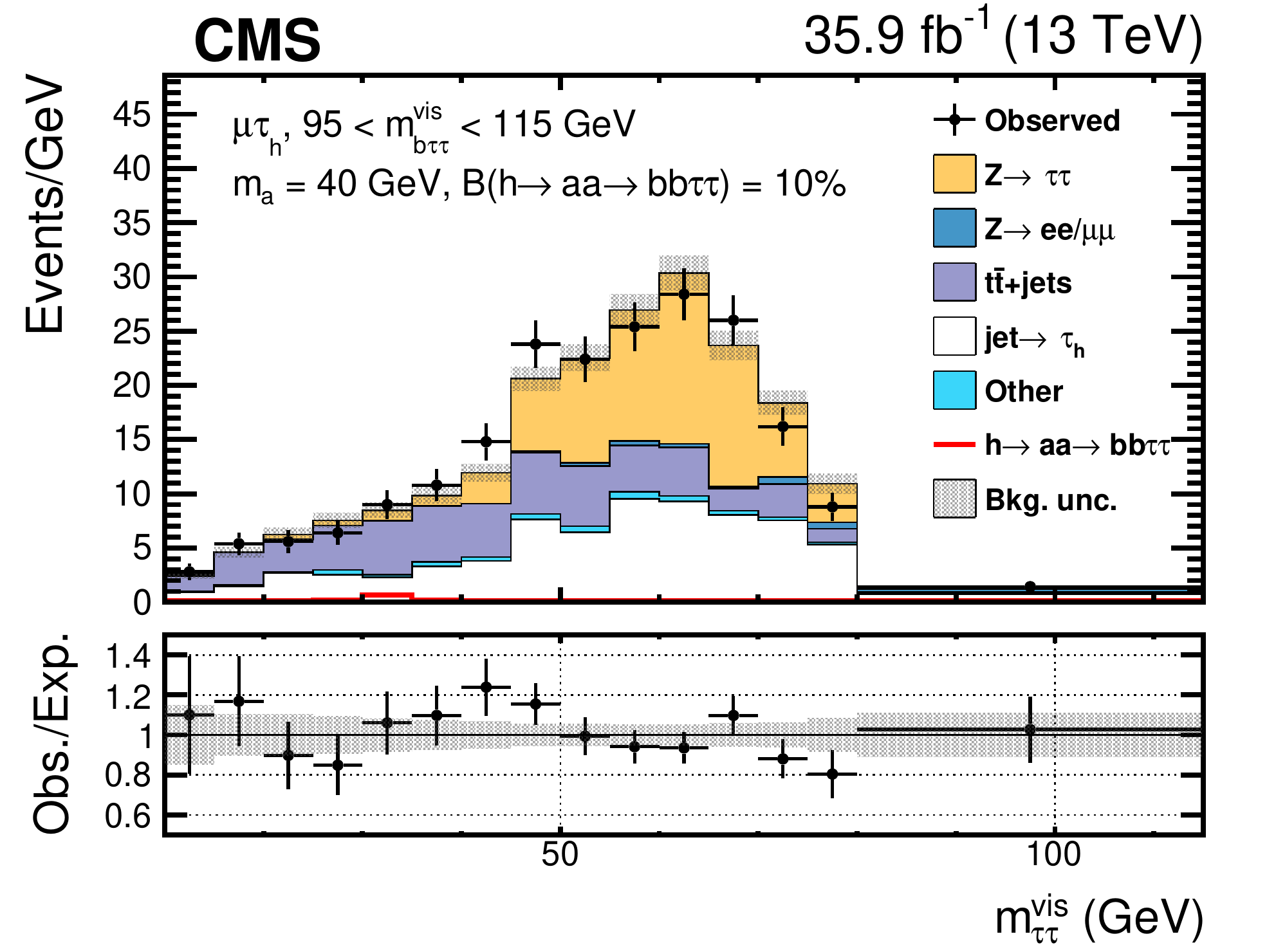}
        \includegraphics[width=0.49\textwidth]{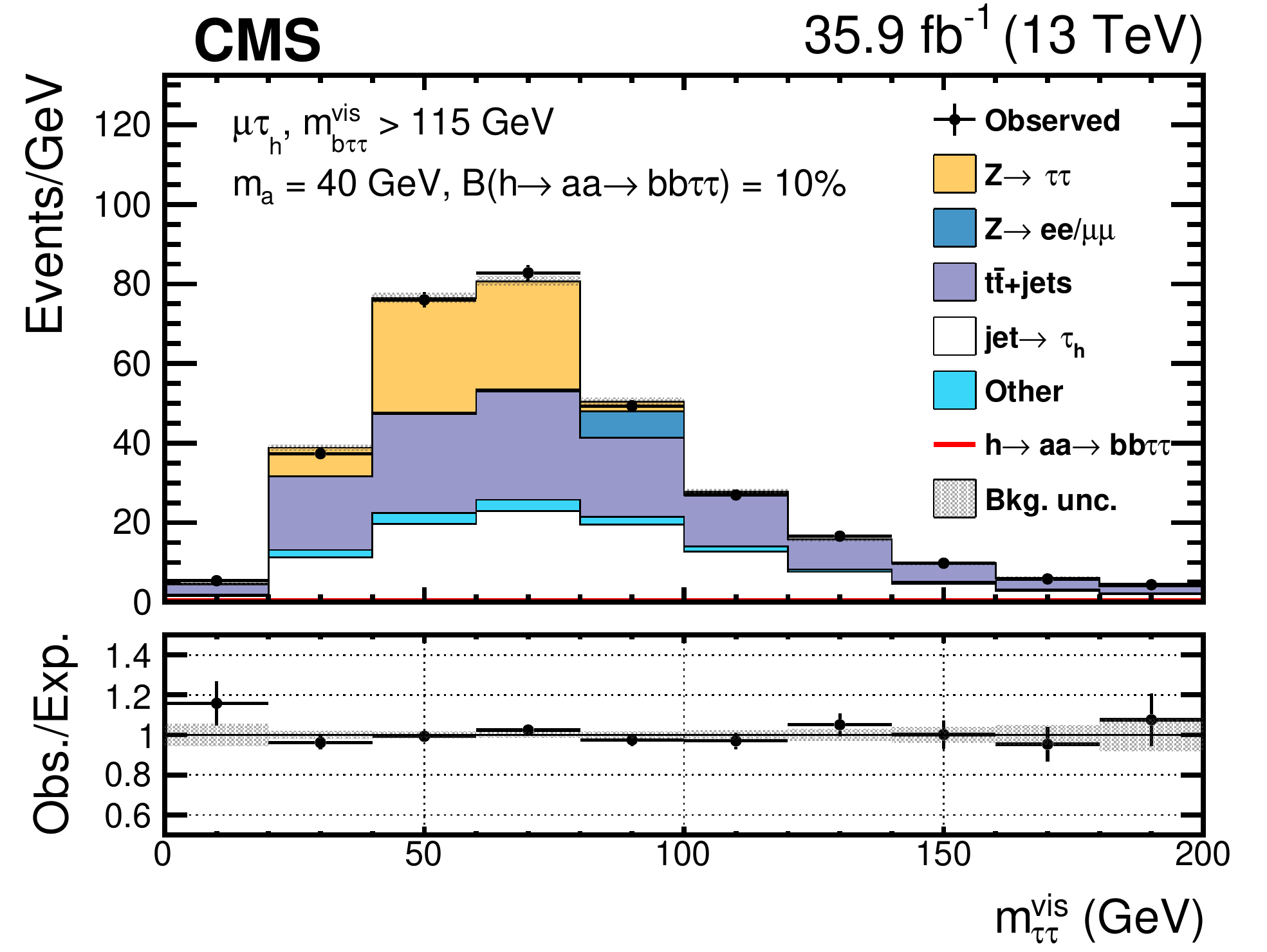}
    \caption{Distributions of $\mvis$ in the four categories of the $\Pgm\tauh$ channel. The ``$\text{jet}\to\tauh$" contribution includes all events with a jet misidentified as a $\tauh$ candidate, whereas the rest of background contributions only include events where the reconstructed $\tauh$ corresponds to a $\tauh$, a muon, or an electron, at the generator level. The ``Other" contribution includes events from single top quark, diboson, and SM Higgs boson processes. The signal histogram corresponds to the SM production cross section for $\cPg\cPg\Ph$, VBF, and V$\Ph$ processes, and assumes $\mathcal{B}(\processbbtt)=10\%$. The normalizations of the predicted background distributions correspond to the result of the global fit.}
    \label{fig:mt_mtt}
\end{figure*}

The $\mvis$ distributions in the different channels and categories are shown in Figs.~\ref{fig:em_mtt}--\ref{fig:mt_mtt}. The binning corresponds to the bins used in the likelihood fit.

\begin{figure*}[hbpt]
\centering
        \includegraphics[width=0.49\textwidth]{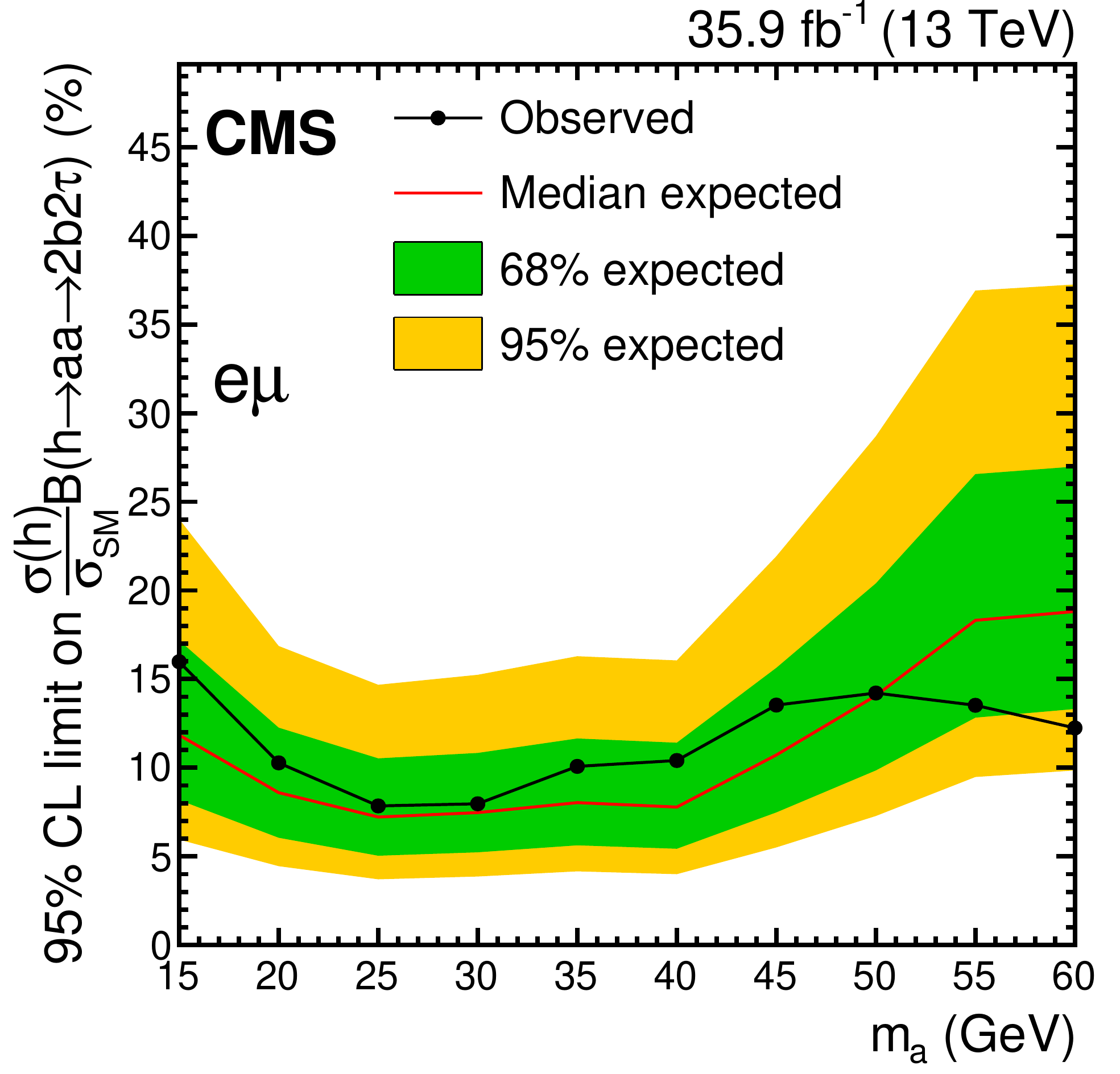}
        \includegraphics[width=0.49\textwidth]{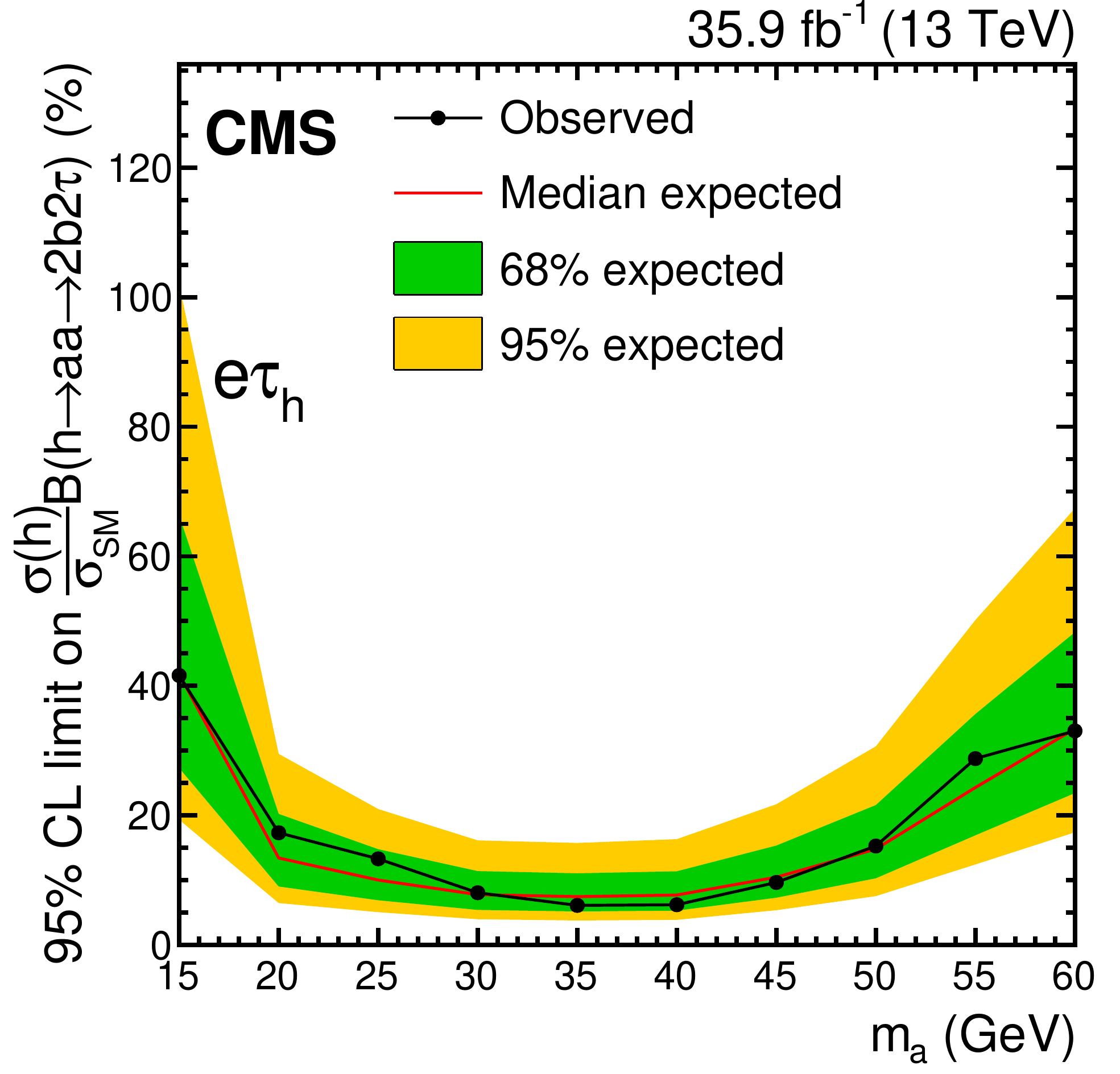}\\
        \includegraphics[width=0.49\textwidth]{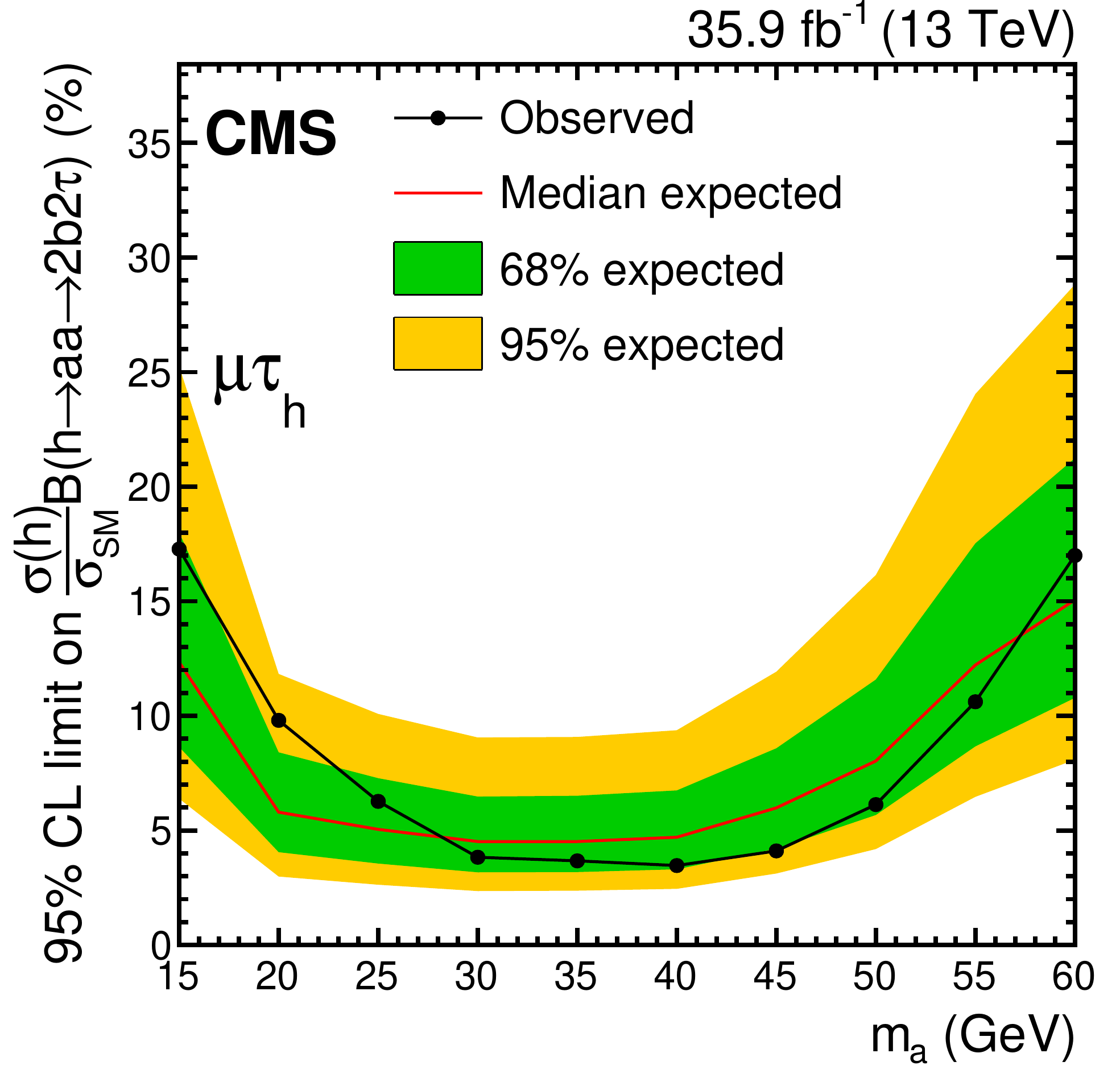}
        \includegraphics[width=0.49\textwidth]{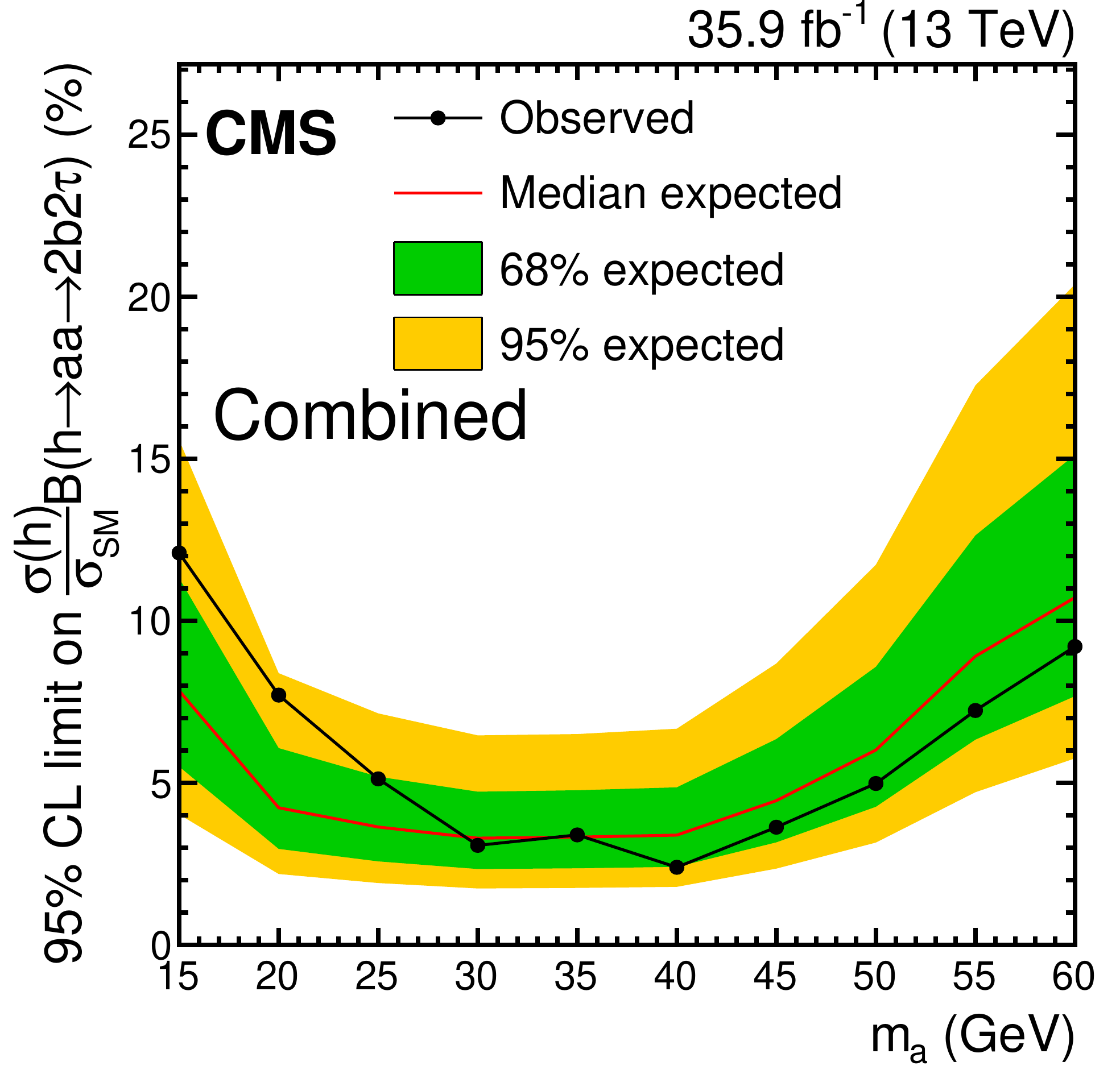}
    \caption{Expected and observed 95\% \CL limits on $(\sigma(\Ph)/\sigma_\mathrm{SM}) \mathcal{B}(\processbbtt)$ in \%. The $\Pe\Pgm$ results are shown in the top left panel, $\Pe\tauh$ in the top right, $\Pgm\tauh$ in the bottom left, and the combination in the bottom right. The inner (green) band and the outer (yellow) band indicate the regions containing 68 and 95\%, respectively, of the distribution of limits expected under the background-only hypothesis.}
    \label{fig:limits}
\end{figure*}

No excess is observed relatively to the SM background prediction. Upper limits at 95\% \CL are set on $(\sigma(\Ph)/\sigma_{\mathrm{SM}}) \mathcal{B}(\processbbtt)$ using the modified frequentist construction \CLs in the asymptotic approximation~\cite{LHC-HCG-Report,Chatrchyan:2012tx,Junk,Read:2002hq,Cowan:2010js}, for pseudoscalar masses between 15 and 60\GeV. In this expression, $\sigma_\mathrm{SM}$ denotes the SM production cross section of the Higgs boson, whereas $\sigma(\Ph)$ is the $\Ph$ production cross section. The limits per channel and for the combination of the three channels are shown in Fig.~\ref{fig:limits}. The most sensitive final state is $\Pgm\tauh$. The sensitivity of the $\Pe\tauh$ and $\Pe\Pgm$ channels is approximately equivalent; the first channel suffers from higher trigger thresholds and lower object identification efficiency than $\Pgm\tauh$, and the second one suffers from a lower branching fraction than $\Pgm\tauh$. At low \ma, the $\Pe\Pgm$ final state has a higher signal acceptance than the other final states, especially $\Pe\tauh$. The limits are more stringent in the intermediate mass range. The low-\ma signals have a lower acceptance because of the overlap of the leptons related to the boost of the pseudoscalar bosons, and of the typically softer lepton and {\cPqb} jet \pt spectra. The high \ma signals lie in a region where more backgrounds contribute, leading also to lower sensitivity than in the intermediate mass region. The categories are complementary over the probed mass range, with the low-$\mbtt$ signal regions more sensitive for heavy resonances, and the high-$\mbtt$ signal regions for light resonances.

\begin{figure}[hbt!]
\centering
        \includegraphics[width=0.49\textwidth]{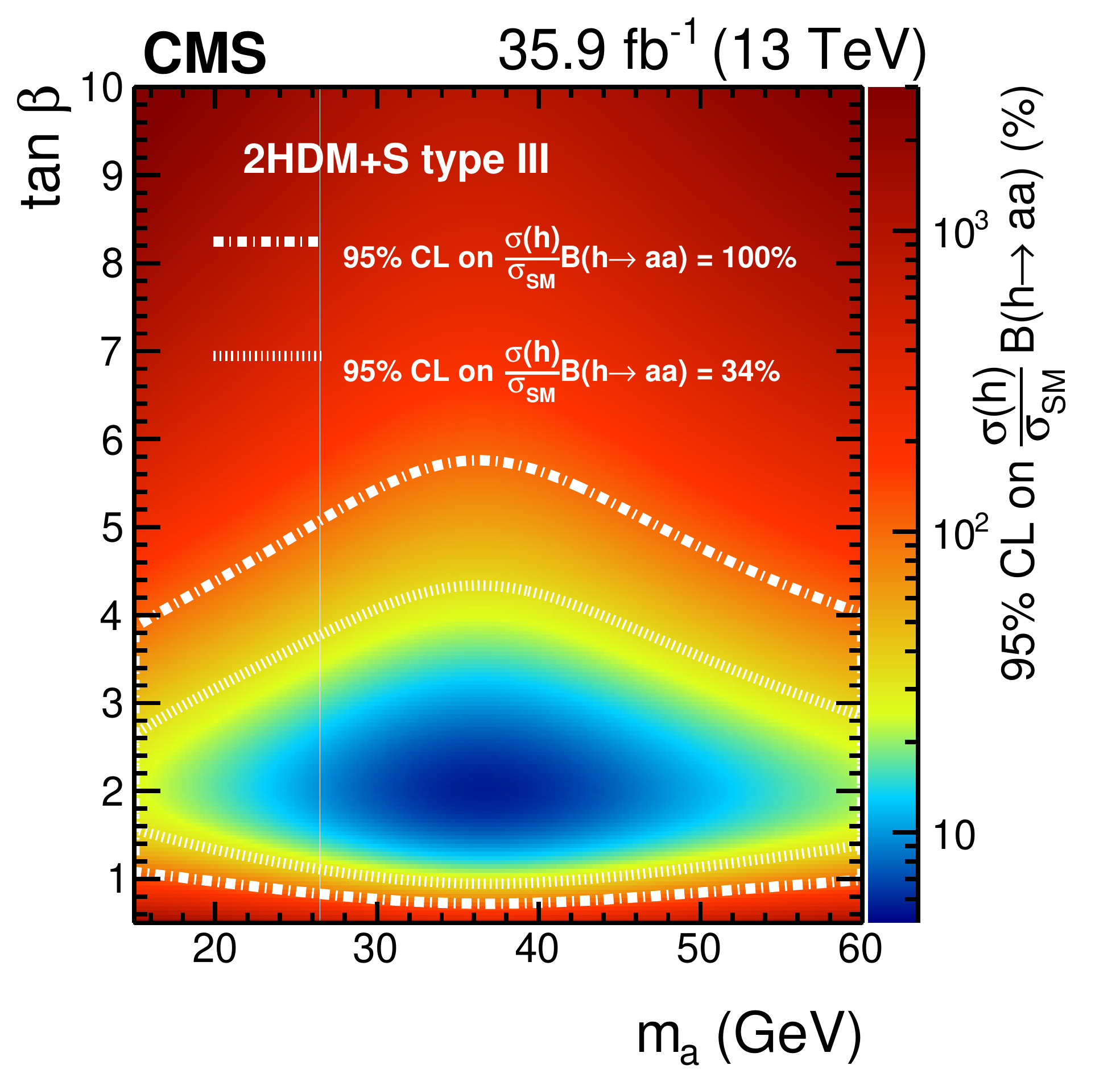}
        \includegraphics[width=0.49\textwidth]{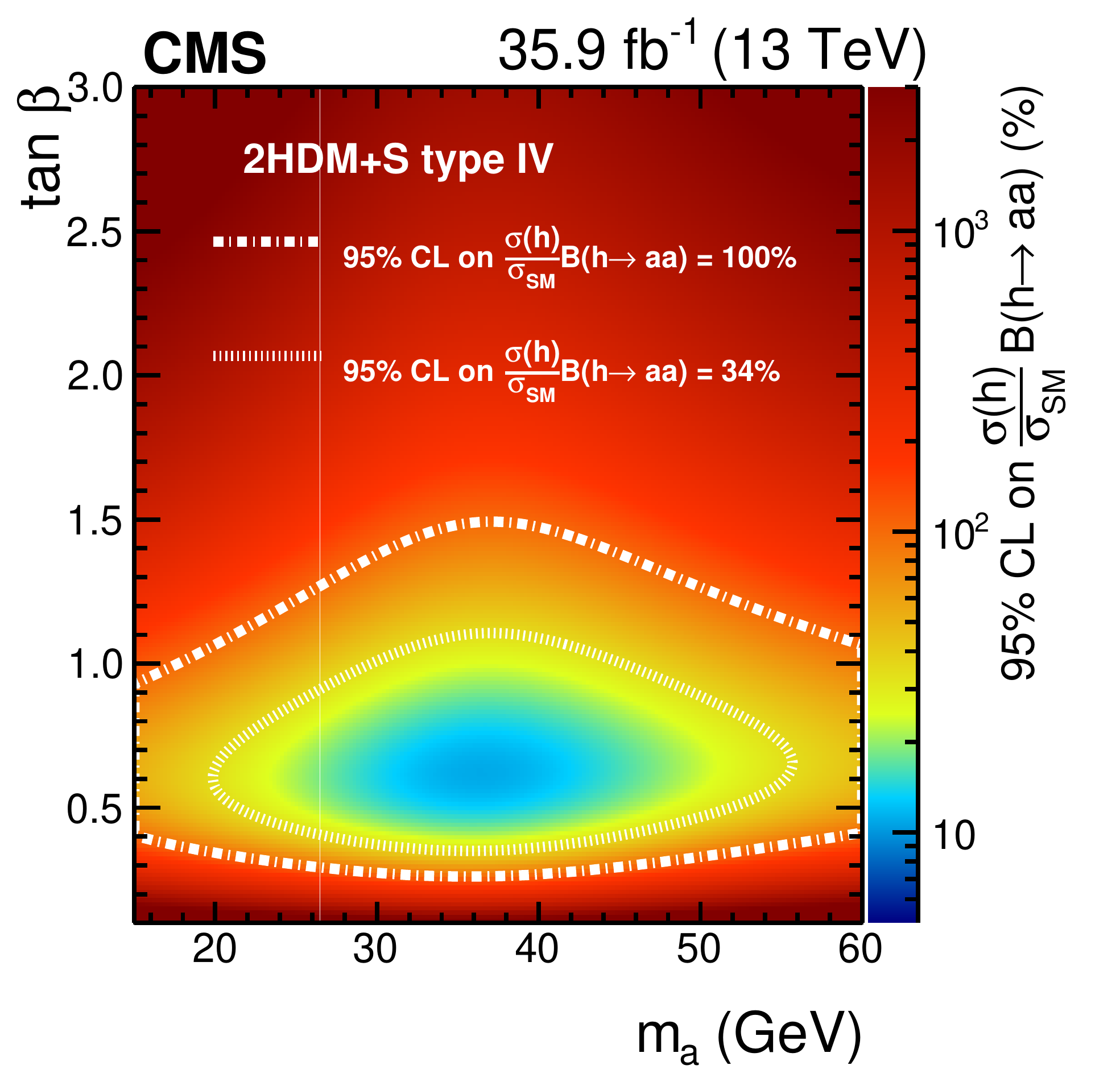}
    \caption{Observed 95\% \CL limits on $(\sigma(\Ph)/\sigma_\mathrm{SM})\mathcal{B}(\Ph\to\Pa\Pa)$ in 2HDM+S of type III (\cmsLeft), and type IV (\cmsRight). The contours corresponding to a 95\% \CL exclusion of $(\sigma(\Ph)/\sigma_\mathrm{SM})\mathcal{B}(\Ph\to \Pa\Pa)=1.00$ and 0.34 are drawn with dashed lines. The number 34\% corresponds to the limit on the branching fraction of the Higgs boson to beyond-the-SM particles at the 95\% \CL obtained with data collected at center-of-mass energies of 7 and 8\TeV by the ATLAS and CMS experiments~\cite{Khachatryan:2016vau}. }
    \label{fig:limits3}
\end{figure}

\begin{figure*}[hbpt]
\centering
        \includegraphics[width=0.70\textwidth]{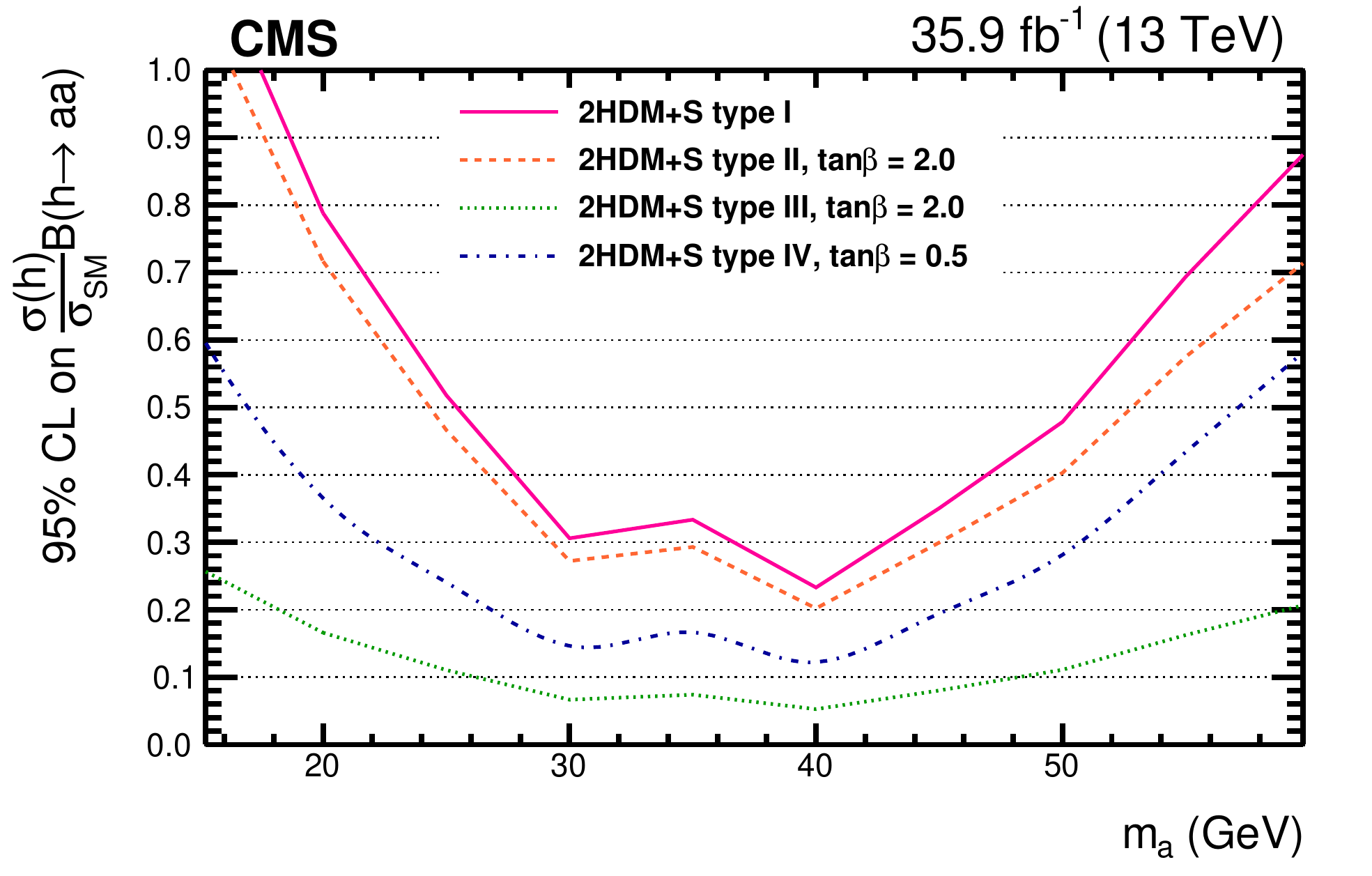}
    \caption{Observed 95\% \CL limits on $(\sigma(\Ph)/\sigma_\mathrm{SM})\mathcal{B}(\Ph\to\Pa\Pa)$ for various 2HDM+S types. The limit in type I 2HDM+S does not depend on $\tan\beta$.}
    \label{fig:limits4}
\end{figure*}

The combined limit at intermediate mass is as low as 3\% on $\mathcal{B}(\processbbtt)$, assuming the SM production cross section and mechanisms for the Higgs boson, and is up to 12\% for the lowest mass point $\ma=15\GeV$. Computing the branching fractions of the light pseudoscalar to SM particles~\cite{PhysRevD.90.075004,anatomy}, this translates to limits on $(\sigma(\Ph)/\sigma_\mathrm{SM})\mathcal{B}(\Ph\to\Pa\Pa)$ of about 20\% in 2HDM+S type II---including the NMSSM---with $\tan\beta>1$ for $\ma=40\GeV$. This improves by more than one order of magnitude previous limits on $\mathcal{B}(\Ph\to\Pa\Pa)$ obtained in the $2\Pgm2\Pgt$ final state by CMS for $15<\ma<25\GeV$~\cite{Khachatryan:2017mnf,Sirunyan:2018mbx}, and by up to a factor five those obtained in the $2\Pgm2\cPqb$ final state by CMS for $25<\ma<60\GeV$~\cite{Khachatryan:2017mnf}.  In the scenario with the highest branching fraction, 2HDM+S type III with $\tan\beta=2$, the expected limit is as low as 6\% at intermediate \ma.  Figure~\ref{fig:limits3} shows the observed limits at 95\% \CL on $(\sigma(\Ph)/\sigma_\mathrm{SM})\mathcal{B}(\Ph\to \Pa\Pa)$ as a function of \ma and $\tan\beta$ for type III and type IV 2HDM+S, for which there is a strong dependence with $\tan\beta$. Figure~\ref{fig:limits4} shows the observed limits at 95\% \CL on $(\sigma(\Ph)/\sigma_\mathrm{SM})\mathcal{B}(\Ph\to\Pa\Pa)$ for a few scenarios of 2HDM+S, assuming the branching fractions of the light pseudoscalar to SM particles computed using Refs.~\cite{PhysRevD.90.075004,anatomy}. The limit shown for type II 2HDM+S is approximately valid for any value of $\tan\beta>1$, and that for type I 2HDM+S does not depend on $\tan\beta$. In the \ma range considered in the analysis, the branching fraction $\mathcal{B}(\Pa\Pa\to \cPqb\cPqb\Pgt\Pgt)$ ranges between 0.10 and 0.11 in type I 2HDM+S, between 0.11 and 0.13 for $\tan\beta=2$ in type II 2HDM+S, between 0.44 and 0.46 for $\tan\beta=2$ in type III 2HDM+S, and between 0.16 and 0.21 for $\tan\beta=0.5$ in type IV 2HDM+S.

\section{Summary}
\label{sec:summary}

The first search for exotic decays of the Higgs boson to pairs of light bosons with two {\cPqb} quark jets and two $\Pgt$ leptons in the final state has been performed with 35.9\fbinv of data collected at 13\TeV center-of-mass energy in 2016. This decay channel has a large branching fraction in many models where the couplings to fermions are proportional to the fermion mass, and can be triggered with high efficiency in the dominant gluon fusion production mode because of the presence of light leptons from leptonic $\Pgt$ decays. No excess of events is found on top of the expected standard model background for masses of the light boson, \ma, between 15 and 60\GeV. Upper limits between 3 and 12\% are set on the branching fraction $\mathcal{B}(\processbbtt)$ assuming the SM production of the Higgs boson. This translates to upper limits on $\mathcal{B}(\Ph\to\Pa\Pa)$ as low as 20\% for $\ma=40\GeV$ in the NMSSM. These results improve by more than one order of magnitude the sensitivity to exotic Higgs boson decays to pairs of light pseudoscalars in the NMSSM from previous CMS results in other final states for $15 < \ma < 25\GeV$, and by a factor up to five for $25 < \ma < 60\GeV$~\cite{Khachatryan:2017mnf,Sirunyan:2018mbx}.

\begin{acknowledgments}
We congratulate our colleagues in the CERN accelerator departments for the excellent performance of the LHC and thank the technical and administrative staffs at CERN and at other CMS institutes for their contributions to the success of the CMS effort. In addition, we gratefully acknowledge the computing centers and personnel of the Worldwide LHC Computing Grid for delivering so effectively the computing infrastructure essential to our analyses. Finally, we acknowledge the enduring support for the construction and operation of the LHC and the CMS detector provided by the following funding agencies: BMWFW and FWF (Austria); FNRS and FWO (Belgium); CNPq, CAPES, FAPERJ, and FAPESP (Brazil); MES (Bulgaria); CERN; CAS, MoST, and NSFC (China); COLCIENCIAS (Colombia); MSES and CSF (Croatia); RPF (Cyprus); SENESCYT (Ecuador); MoER, ERC IUT, and ERDF (Estonia); Academy of Finland, MEC, and HIP (Finland); CEA and CNRS/IN2P3 (France); BMBF, DFG, and HGF (Germany); GSRT (Greece); NKFIA (Hungary); DAE and DST (India); IPM (Iran); SFI (Ireland); INFN (Italy); MSIP and NRF (Republic of Korea); LAS (Lithuania); MOE and UM (Malaysia); BUAP, CINVESTAV, CONACYT, LNS, SEP, and UASLP-FAI (Mexico); MBIE (New Zealand); PAEC (Pakistan); MSHE and NSC (Poland); FCT (Portugal); JINR (Dubna); MON, RosAtom, RAS and RFBR (Russia); MESTD (Serbia); SEIDI, CPAN, PCTI and FEDER (Spain); Swiss Funding Agencies (Switzerland); MST (Taipei); ThEPCenter, IPST, STAR, and NSTDA (Thailand); TUBITAK and TAEK (Turkey); NASU and SFFR (Ukraine); STFC (United Kingdom); DOE and NSF (USA).

\hyphenation{Rachada-pisek} Individuals have received support from the Marie-Curie program and the European Research Council and Horizon 2020 Grant, contract No. 675440 (European Union); the Leventis Foundation; the A. P. Sloan Foundation; the Alexander von Humboldt Foundation; the Belgian Federal Science Policy Office; the Fonds pour la Formation \`a la Recherche dans l'Industrie et dans l'Agriculture (FRIA-Belgium); the Agentschap voor Innovatie door Wetenschap en Technologie (IWT-Belgium); the F.R.S.-FNRS and FWO (Belgium) under the ``Excellence of Science - EOS" - be.h project n. 30820817; the Ministry of Education, Youth and Sports (MEYS) of the Czech Republic; the Lend\"ulet (``Momentum") Programme and the J\'anos Bolyai Research Scholarship of the Hungarian Academy of Sciences, the New National Excellence Program \'UNKP, the NKFIA research grants 123842, 123959, 124845, 124850 and 125105 (Hungary); the Council of Science and Industrial Research, India; the HOMING PLUS program of the Foundation for Polish Science, cofinanced from European Union, Regional Development Fund, the Mobility Plus program of the Ministry of Science and Higher Education, the National Science Center (Poland), contracts Harmonia 2014/14/M/ST2/00428, Opus 2014/13/B/ST2/02543, 2014/15/B/ST2/03998, and 2015/19/B/ST2/02861, Sonata-bis 2012/07/E/ST2/01406; the National Priorities Research Program by Qatar National Research Fund; the Programa Estatal de Fomento de la Investigaci{\'o}n Cient{\'i}fica y T{\'e}cnica de Excelencia Mar\'{\i}a de Maeztu, grant MDM-2015-0509 and the Programa Severo Ochoa del Principado de Asturias; the Thalis and Aristeia programs cofinanced by EU-ESF and the Greek NSRF; the Rachadapisek Sompot Fund for Postdoctoral Fellowship, Chulalongkorn University and the Chulalongkorn Academic into Its 2nd Century Project Advancement Project (Thailand); the Welch Foundation, contract C-1845; and the Weston Havens Foundation (USA). \end{acknowledgments}

\bibliography{auto_generated}

\cleardoublepage \appendix\section{The CMS Collaboration \label{app:collab}}\begin{sloppypar}\hyphenpenalty=5000\widowpenalty=500\clubpenalty=5000\vskip\cmsinstskip
\textbf{Yerevan Physics Institute, Yerevan, Armenia}\\*[0pt]
A.M.~Sirunyan, A.~Tumasyan
\vskip\cmsinstskip
\textbf{Institut f\"{u}r Hochenergiephysik, Wien, Austria}\\*[0pt]
W.~Adam, F.~Ambrogi, E.~Asilar, T.~Bergauer, J.~Brandstetter, E.~Brondolin, M.~Dragicevic, J.~Er\"{o}, A.~Escalante~Del~Valle, M.~Flechl, R.~Fr\"{u}hwirth\cmsAuthorMark{1}, V.M.~Ghete, J.~Hrubec, M.~Jeitler\cmsAuthorMark{1}, N.~Krammer, I.~Kr\"{a}tschmer, D.~Liko, T.~Madlener, I.~Mikulec, N.~Rad, H.~Rohringer, J.~Schieck\cmsAuthorMark{1}, R.~Sch\"{o}fbeck, M.~Spanring, D.~Spitzbart, A.~Taurok, W.~Waltenberger, J.~Wittmann, C.-E.~Wulz\cmsAuthorMark{1}, M.~Zarucki
% ...
\vskip\cmsinstskip
% ...
1:  Also at Vienna University of Technology, Vienna, Austria\\
% ...
\end{sloppypar}
\end{document}